\pdfoutput=1
\documentclass[superscriptaddress,prd,twocolumn,nofootinbib]{revtex4-2}
\usepackage[colorlinks=true, pdfstartview=FitV, linkcolor=red, citecolor=blue,
urlcolor=blue]{hyperref}
\usepackage[dvipdfmx]{graphicx}
\usepackage{amsmath,amssymb}
\usepackage{color}
\usepackage{slashed}
\usepackage[normalem]{ulem}
\renewcommand\a{\alpha}
\renewcommand\b{\beta}
\renewcommand\d{\delta}

\renewcommand\l{\lambda}
\renewcommand\r{\rho}

\renewcommand\t{\tau}

\renewcommand\j{\psi}

\newcommand\g{\gamma}

\newcommand\m{\mu}
\newcommand\n{\nu}

\newcommand\p{\pi}
\newcommand\h{\theta}
\newcommand\s{\sigma}
\newcommand\f{\phi}

\newcommand\ve{\varepsilon}

\newcommand\vf{\varphi}
\renewcommand\L{\Lambda}

\renewcommand\O{\Omega}

\newcommand\D{\Delta}
\newcommand\G{\Gamma}
\newcommand\F{\Phi}

\newcommand{\diag}{{\rm{diag}}}
\newcommand{\eff}{{\rm{eff}}}
\newcommand{\B}{{\rm{B}}}

\newcommand{\Tr}{{\rm Tr}}

\newcommand{\sgn}{{\rm sgn}}

\newcommand{\Lag}{\mathcal{L}}

\newcommand{\llk}{\lambda_{l,k}}

\newcommand{\ud}{\mathrm{d}}
\newcommand{\ue}{\mathrm{e}}
\newcommand{\oh}{\frac{1}{2}}

\newcommand\px{\partial_x}
\newcommand\py{\partial_y}

\newcommand\prm{^\prime}
\newcommand\sprm{^{*\prime}}
\newcommand{\fig}[1]{Fig.~\ref{#1}}

\newcommand\ra{\rightarrow}

\renewcommand{\part}{{\rm part}}
\newcommand{\rout}{\bgroup \color{red} \ULdepth=-.5ex \ULset}

\begin{abstract}
We revisit the condensation scenario of charged pions in external magnetic field and rotation, which was first considered by Y. Liu and I. Zahed.
Based on the Ginzburg-Landau analysis of the Nambu--Jona-Lasinio model, we find that the charged-pion condensation takes place only when both a strong coupling constant and negatively large baryon chemical potential are applied.
Besides, our numerical calculation shows that the chiral restoration induced by the interplay between magnetic field and rotation (i.e., the rotational magnetic inhibition) interrupts the formation of the charged-pion condensate.
This suggests that the analysis of such condensation requires a careful treatment of the inner structure of pions, which was not taken into account before. We also discuss the underlying physical mechanism of our finding and the indication of charged-rho condensation.
\end{abstract}
\begin{document}

\title{Do charged-pions condense in a magnetic field with rotation?}

\author{Hao-Lei Chen}
\email{hlchen15@fudan.edu.cn}
\affiliation{Key Laboratory of Nuclear Physics and Ion-beam Application (MOE), Fudan University, Shanghai 200433, China}
\affiliation{Shanghai Research Center for Theoretical Nuclear Physics, NSFC and Fudan University, Shanghai 200438, China}

\author{Xu-Guang Huang}
\email{huangxuguang@fudan.edu.cn}
\affiliation{Physics Department and Center for Particle Physics and Field Theory, Fudan University, Shanghai 200438, China}
\affiliation{Key Laboratory of Nuclear Physics and Ion-beam Application (MOE), Fudan University, Shanghai 200433, China}
\affiliation{Shanghai Research Center for Theoretical Nuclear Physics, NSFC and Fudan University, Shanghai 200438, China}

\author{Kazuya Mameda}
\email{k.mameda@rs.tus.ac.jp}
\affiliation{Physics Department and Center for Particle Physics and Field Theory, Fudan University, Shanghai 200438, China}
\affiliation{Department of Physics, Tokyo University of Science, Tokyo 162-8601, Japan}
\affiliation{RIKEN iTHEMS, RIKEN, Wako 351-0198, Japan}

%\pacs{}
\maketitle

\section{Introduction}
In the last few decades, quark matter under a magnetic background has attracted many attentions in a wide range of physical systems from the early universe to the neutron stars and heavy ion collision experiments.
One of the active research topics is the anomalous chiral transport phenomena caused by the quantum anomaly in a magnetic field, for example, the chiral magnetic effect (CME)~\cite{Kharzeev:2007jp,Fukushima:2008xe} and its cousins.
Their experimental searches are the frontiers of the recent heavy-ion collision physics~\cite{Huang:2015oca,Hattori:2016emy,Kharzeev:2015znc}.
The magnetic response of quark matter also strongly affects the phase structure of quantum chromodynamics (QCD).
A lot of interesting phenomena have been discussed in this context, e.g., the magnetic catalysis of chiral condensate~\cite{Gusynin:1994re,Gusynin:1995nb,Miransky:2015ava}, inverse magnetic catalysis at finite temperature~\cite{Bali:2011qj,Bali:2012zg} and density~\cite{Preis:2010cq}, and others~\cite{Fukushima:2007fc,Ferrer:2007iw,Son:2007ny,Chernodub:2011mc,Hidaka:2012mz,Sinha:2013dfa,Cao:2015cka,Brauner:2016pko,Hattori:2017qio,Ozaki:2015sya,Chen:2017xrj}.

Additionally, various properties of rotating relativistic matter has been actively discussed, motivated by extremely strong fluid vorticity found in heavy-ion collisions~\cite{Deng:2016gyh,Jiang:2016woz,Wei:2018zfb} via the spin polarization of spinful particles~\cite{Liang:2004ph,STAR:2017ckg}.
Such a vorticity can generate the parity violating current called the chiral vortical effect (CVE)~\cite{Vilenkin:1979ui,Vilenkin:1980zv,Son:2009tf,Liu:2018xip}, which is analogous to the CME.
It is also revealed that rotation affects even the phase structure of QCD~\cite{Chen:2015hfc,Jiang:2016wvv,Ebihara:2016fwa,Chernodub:2016kxh,Huang:2017pqe,Chernodub:2017ref,Chernodub:2017mvp,Liu:2017spl,Wang:2018zrn,Zhang:2018ome,Wang:2018sur}.
In a uniformly rotating system at finite temperature and/or density, angular velocity plays a role as an effective chemical potential, and thus suppresses the spin-0 pairing of quarks~\cite{Jiang:2016wvv}.
On the other hand, it is confirmed that such a rotational effect on thermodynamics is invisible at zero temperature and density~\cite{Ebihara:2016fwa,Chernodub:2016kxh,Chernodub:2017ref}.
In Refs.~\cite{Wang:2018zrn,Wang:2019nhd}, the authors showed that under a nonuniform rotation the ground state exhibits a vortex structure under sufficiently rapid rotation.

On top of these, a remarkable property of rotating matter is that combined with magnetic field, rotation creates more fruitful QCD phase structure.
In Ref.~\cite{Chen:2015hfc}, Fukushima and the present authors found, from the Nambu--Jona-Lasinio (NJL) model analysis, that the chiral condensate decreases as magnetic field increasing, and eventually the chiral symmetry is restored.
This novel phenomenon was named the ``rotational magnetic inhibition", as an analogy to the magnetic inhibition phenomenon in the finite density system.
Furthermore, it is argued in Refs.~\cite{Liu:2017spl,Liu:2017zhl} that the interplay between magnetic field and rotation can induce a charged-pion condensation.
The essential idea is that such a combined effect leads to an energy splitting between $\pi^+$ and $\pi^-$, if pions are taken into account as {\it point-like} particles (see Section~\ref{sec:gene}).
Namely, an effectively isospin chemical potential is induced.
As a result, a charged-pion Bose-Einstein condensation (BEC) takes place, similarly to that induced in a finite isospin density system~\cite{Son:2000xc,He:2005nk,Sun:2007fc,He:2006tn}.

In the above argument of the charged-pion condensate, however, it is unclear whether the point-particle picture is safely admitted.
Indeed, it would be expected that the inner structure is important, in the following two senses.
First, due to the rotational magnetic inhibition, magnetic field and rotation would suppress not only the chiral condensate but also the charged-pion condensate.
Second, since both magnetic field and rotation tend to align the angular momenta of the paired quark and antiquark, the magnetized and rotating quark system energetically disfavors the condensation of charged-pion, which is a spin-0 composite state.
%Therefore, we should carefully compare the interplay between magnetic field and rotation on
%the spectrum splitting by rotation and magnetic field, and the rotational magnetic inhibition.

The main purpose of this paper is to quantify how the quark dynamics affects the charged-pion condensation induced in magnetic field and rotation.
For this reason, we perform the Ginzburg-Landau analysis of the $\pi^\pm$ fields in the two-flavor NJL model.
We find that within this analysis, the charged-pion BEC is not observed for a small coupling strength.
However, if a negatively high baryon chemical potential is present, the charged-pion condensation can take place for a strong coupling constant.
We also discuss the underlying physical meaning of these results.

%%%%%%%%%%%%%%%%%%%%%%%%%%%%%%%%%%%%%%%%%%%%%%%%%%%%%%%%%%%%%%%%%%%%%%%
\section {General Arguments}\label{sec:gene}
%%%%%%%%%%%%%%%%%%%%%%%%%%%%%%%%%%%%%%%%%%%%%%%%%%%%%%%%%%%%%%%%%%%%%%%
First let us review the condensation mechanism of the {\it point-like} particle $\pi^\pm$~\cite{Liu:2017spl,Liu:2017zhl}.
We consider the magnetized and rotating cylindrical system with a magnetic field $\vec{B}=B\hat{z}$ , angular velocity $\vec{\Omega}= \Omega\hat{z}$, and the transverse system radius $R$.
In this paper, we always assume $e>0$ and $B>0$ and $\Omega>0$.
The dispersion relations of the pions are given by
\begin{equation}
	\label{eq:Epi}
   E_{\pi^\pm}=\sqrt{eB(2n+1)+p_z^2+m_\pi^2}\mp\Omega l,
 \end{equation}
where $n$ ($l$) is the quantum number of the Landau levels (angular momentum).
The above dispersion implies that $\pi^\pm$ is affected by an effective chemical potential $\pm \Omega l$.
Therefore the lowest Landau level (LLL) $\pi^+$ forms a Bose condensate when $\mu_N=N\Omega$ (with $N$ being the magnetic flux, and thus the maximum value of $l$) exceeds $m_0=\sqrt{m_\pi^2+eB}$.
Besides, as $\Omega$ increased, a higher Landau level $\pi^+$ may Bose-condense; see Refs.~\cite{Liu:2017spl,Liu:2017zhl} for more details.
The condensate is possibly inhomogeneous as discussed in ~\cite{Guo:2021gbz}.
We note that the condition $\O\leqslant 1/R\ll\sqrt{eB}$ is implicitly assumed:
the first inequality is due to the causality, and the second one is to make the Landau quantization sensible~\cite{Chen:2015hfc}.

This picture cannot be applicable, however, when pions are regarded as composite spin-0 particles of quarks and antiquarks (especially, in a large magnetic field $eB > \L_{\rm QCD}$).
Let us consider $\p^+$, which comprise a $u$ quark and a $\bar{d}$ antiquark with their angular momenta (both the spin and the orbital one) antiparallel to each other.
Since the magnetic field and rotation would tend to align the angular momenta of $u$ and $\bar{d}$, the $\p^+$ condensate is suppressed.
In addition, if a magnetic field is strong enough, the chiral restoration also takes place via the rotational magnetic inhibition~\cite{Chen:2015hfc} with an effective baryon chemical potential $\sim \O N$%
~\footnote{The appearance of the baryon density contains a contribution from chiral anomaly such that $n_\B\sim \O B/(4\p^2)$~\cite{Hattori:2016njk,Ebihara:2016fwa}, which persists even in low-energy effective theory due to anomaly matching. Thus, even if we disregard the inner quark structure of pions, a finite baryon density still exists.%
}.
It is hence necessary to discuss the charged-pion condensation in the language of quarks.

In the rest of this article, we will use an NJL model to study the above two competing effects.
The NJL model lacks the confinement and has no scale like $\L_{\rm QCD}$ to separate the hadronic degrees of freedom from the quarks.
The results from the NJL model may differ from that of QCD, but usually provide useful insights to QCD physics, especially those related to symmetry breaking.
The $\p^+$ condensate in NJL model is considered as $\langle \bar{d}\mathrm{i}\g^5 u\rangle$, which can smoothly change from the condensates of loosely correlated pairs to tightly bound bosons upon tuning the coupling constant or densities, a.k.a., BCS (Bardeen-Cooper-Schrieffer)-BEC crossover.
For a wide range of the coupling constant, we do not observe qualitatively different results, so in the following we present our results for a few fixed coupling constant.

We note that the charged-pion condensation also triggers an electric superconductivity, which always obstructs the penetration of the magnetic field (the Meissner effect%
~\footnote{
	For a rotating superconductor, the superfluid velocity is entangled with the magnetic field, which makes the discussion of the Meissner effect more involved.
	But the main implication to the charged-pion condensation is unchanged.%
	}).
As a result, there are two possibilities;
the magnetic field is repelled from the charged-pion condensate, and thus it cannot induce pion condensation at all;
or the magnetic field penetrates into the condensate from, e.g. the magnetic vortices if it is a type-II superconductor, to turn the system very inhomogeneous.
These issues will not be discussed in the following.
%%%%%%%%%%%%%%%%%%%%%%%%%%%%%%%%%%%%%%%%%%%%%%%%%%%%%%%%%%%%%%%%%%%%%%%
\section {Dirac Equation}\label{sec:Dirac}
%%%%%%%%%%%%%%%%%%%%%%%%%%%%%%%%%%%%%%%%%%%%%%%%%%%%%%%%%%%%%%%%%%%%%%%
We start from the Dirac equation in rotating frame with a magnetic field $\vec B$.
The frame is rotating with a constant angular velocity $\vec \Omega=\Omega\hat z$, which can be described by the metric tensor
  \begin{equation}
  g_{\m\n}=
	\begin{pmatrix}
		1-r^2\O^2 &\O y& -\O x& 0 \\
       \O y & -1 & 0 & 0 \\
       -\O x & 0 & -1 & 0 \\
       0 & 0 & 0 & -1 \\	
	\end{pmatrix}
\end{equation}
with $r^2 = x^2 + y^2$.
The Dirac equation in such a frame is
\begin{equation}\label{eq:cDriac}
  \Bigl[
  	\mathrm{i}{\gamma}^\mu (\partial_\mu+\mathrm{i}qA_\mu+\Gamma_\mu)-m+\mu\gamma^0
  \Bigr] \psi(x)=0\,,
\end{equation}
where $A_\mu$ is the background gauge field in the rotating frame, $q$ the charge of the Dirac fermion, $\m$ the fermion chemical potential.
The spin connection $\Gamma_\mu$ is defined by
\begin{equation}
  \begin{split}
    \Gamma_\mu&=-\frac{\mathrm{i}}{4}\omega_{\mu ab}\sigma^{ab}\,,\\
    \omega_{\mu ab}&=g_{\a\b}e^\a_a(\partial_\mu e^\b_b+\G^\b_{\n\m}e^\n_b)\,,\\
    \sigma^{ab}&=\frac{\mathrm{i}}{2}[\g^a,\g^b]\,.
  \end{split}
\end{equation}
The Greek and the Latin indices stand for the curved and local Lorentz coordinates, respectively. The vierbein field $e^\mu_a$ is chosen to be
\begin{equation}
  e^t_0=e^x_1=e^y_2=e^z_3=1,\quad e^x_0=y\Omega,\quad e^y_0=-x\Omega,
\end{equation}
and other components are zero.

We now consider the situation that a constant magnetic field is set to be along the rotating axis.
In the non-rotating Minkowski spacetime with the coordinate $(t',x',y',z')$, the magnetic field is along the $z'$-axis, i.e., $\vec B=B \hat z'$.
We adopt the symmetric gauge $A_a^\prime=(0,By'/2,-Bx'/2,0)$, then get the gauge potential in the rotating frame by a coordinate transformation: $A_\mu(x)=(-\oh B\O r^2,\oh By,-\oh Bx,0)$.
Thus, the Dirac equation is written as
\begin{equation}
\begin{split}
    \mathrm{i}\partial_t\j&=(\hat H_D-\Omega \hat J_z-\mu)\j\,,\\
  \hat H_D&=-\mathrm{i}\gamma^0 \gamma^1\left(\partial_x+\mathrm{i}qB{y\over2}\right)-i\gamma^0 \gamma^2\left(\partial_y-\mathrm{i}qB{x\over2}\right)\\
  &\;\;\;\;\,-\mathrm{i}\gamma^0 \gamma^3\partial_z+m\gamma^0\,,\\
  \hat J_z&=\mathrm{i}y\px-\mathrm{i} x\py+\frac{\s_z}{2}=-\mathrm{i}\partial_\theta+\frac{\s_z}{2}\,.
\end{split}
\end{equation}

A uniformly rotating system must be finite-sized.
We consider a cylindrical system with a radius $R$ satisfying the causality constraint $\Omega R<1$, and suppose that an appropriate boundary condition is imposed.
%The boundary condition is given by (i.e., a weak version of no-flux boundary condition)~\cite{Chen:2017xrj}
%%(we omit $l$, $k$ indices of the wave function here for simplicity)
For $qB>0$, the solutions can be written as
\begin{equation}
  \psi^{(a)}_s
  =\frac{1}{\sqrt{\pi R^2 N_{l,k}^2}}
	\frac{1}{\sqrt{2 \varepsilon}}
  	\ue^{-\mathrm{i}(a\ve-\Omega j-\mu) t}\Psi^{(a)}_s
\end{equation}
with $\ve=\sqrt{2qB\l_{l,k}+p_z^2+m^2}$ and $\l_{l,k}$ being the $k$-th eigenvalue of the radial Hamiltonian for angular momentum quantum number $l$.
Also $a=\pm$ stands for positive or negative frequency solution, $s=\pm$ represents the spin state, and $j=l+1/2$.
Note that $\l_{l,k}$ and $N_{l,k}$ depend on what types of boundary condition is imposed~\cite{Chen:2017xrj}.
The eigenfunctions of $\hat H_D$ with the eigenvalue $a \varepsilon$ reads%
~\footnote{%
Note that the present convention is different from that in Ref.~\cite{Chen:2017xrj}.%
}
\begin{eqnarray}
   \Psi^{(+)}_+&=&\frac{\ue^{\mathrm{i}p_zz}}{\sqrt{\ve+m}}
   \begin{pmatrix}
    	(\ve+m)\f_{l,k} \\
    	0 \\
    	p_z\f_{l,k}\\
    	\mathrm{i}\sqrt{2qB\l_{l,k}}\vf_{l,k}
   \end{pmatrix},\\
  \Psi^{(+)}_-&=&\frac{\ue^{\mathrm{i}p_zz}}{\sqrt{\ve+m}}
   \begin{pmatrix}
   		0\\
       (\ve+m)\vf_{l,k} \\
		-\mathrm{i}\sqrt{2qB\l_{l,k}}\f_{l,k}\\
		-p_z\vf_{l,k}
   \end{pmatrix},\\
 \Psi^{(-)}_+&=&\frac{\ue^{-\mathrm{i}p_zz}}{\sqrt{\ve+m}}
	\begin{pmatrix}
		-p_z\f_{l,k}\\
       \mathrm{i}\sqrt{2qB\l_{l,k}}\vf_{l,k}\\
       -(\ve+m)\f_{l,k} \\
       0
	\end{pmatrix},\\
  \Psi^{(-)}_-&=&\frac{\ue^{-\mathrm{i}p_zz}}{\sqrt{\ve+m}}
  	\begin{pmatrix}
  		-\mathrm{i}\sqrt{2qB\l_{l,k}}\f_{l,k}\\
		p_z\vf_{l,k}\\
		 0\\
       - (\ve+m)\vf_{l,k}
  	\end{pmatrix}.
\end{eqnarray}
Here we have introduced
\begin{equation}
\begin{split}
  \phi_{l,k}
  &=\ue^{il\h}\Phi_{l}(\l_{l,k},\tfrac{1}{2} qBr^2)\,,\\
  \varphi_{l,k}
  &={\rm sgn}(j)\ue^{i(l+1)\h}\Phi_{l+1}(\l_{l,k}-1,\tfrac{1}{2} qBr^2)\,,
\end{split}
\end{equation}
where
\begin{equation}
\begin{split}
  \Phi_{l}(\l,x)=&\frac{1}{|l|!}\Bigg[\frac{\G(\l+l+1)}{\G(\l+1)} \Bigg]^{\frac{{\rm sgn}(j)}{2}}x^{\frac{|l|}{2}}\ue^{-x/2}\\
  &\times{}_1F_1(-\l+\frac{|l|-l}{2},|l|+1,x)
  \end{split}
\end{equation}
with $_1F_1$ the Kummer's function of the first kind. If $\lambda$ is integer, i.e. $R\to \infty$, we have
\begin{equation}
   \Phi_{l}(\l,x)
   = \biggl[\frac{\lambda !}{(\lambda+l)!} \biggr]^{\frac{{\rm sgn}(j)}{2}}x^{\frac{|l|}{2}}\ue^{-x/2}L^{|l|}_{\l-(|l|-l)/2}(x)
\end{equation}
with $L^{l}_{n}$ the Laguerre polynomials.
We can construct the propagator from the solutions,
\begin{equation}
  \begin{split}
  \label{eq:SFu}
  S(x,x\prm)&=
  		\mathrm{i}\int\frac{\ud p_0\ud p_z}{(2\pi)^2}
  		\sum_{l=-\infty}^\infty\sum_{k=1}^\infty\frac{1}{\pi R^2N^2_{lk}}\\
	&\quad
		\times\frac{\ue^{-\mathrm{i} (p_0-\Omega j -\mu)\D t+\mathrm{i}p_z\D z} }{p_0^2- \varepsilon^2+\mathrm{i}\epsilon}
		 \mathcal S(p_0),
  \end{split}
\end{equation}
with $\D x=x-x'$ and
\begin{eqnarray}
 & \mathcal S(p_0) =
  \begin{pmatrix}
  	 \mathcal M_+ & \mathcal N \\
     -\mathcal N & \mathcal M_-
  \end{pmatrix},\\
 & \mathcal M_\pm = (\pm p_0+m)
  	\begin{pmatrix}
  		\f\f\sprm & 0 \\
    	0 & \vf\vf\sprm
  	\end{pmatrix},\\
 &   \mathcal N =
	\begin{pmatrix}
		-p_z\f\f\sprm & \mathrm{i}\sqrt{2qB\llk}\f\vf\sprm \\
		-\mathrm{i}\sqrt{2qB\llk}\vf\f\sprm & p_z\vf\vf\sprm
	\end{pmatrix}.
\end{eqnarray}
Here we have abbreviated $\f_{l,k}(x)$ by $\f$, $\f^*_{l,k}(x\prm)$ by $\f\sprm$, and so on.

%For $qB<0$, the solutions $\tilde\psi^{(a)}_s$ can be obtained by charge conjugation,
%\begin{equation}
%  \tilde\psi^{(a)}_s=-i\g^2\psi^{(-a)*}_{-s}.
%\end{equation}
The propagator for $qB<0$ is obtained from Eq.~\eqref{eq:SFu}.
The difference between the propagators for $qB>0$ and $qB<0$ appears only in the sign of angular momentum;
when the sign of $qB$ changes, a charged particle moves to an opposite direction and its spin is also flipped.
For this reason, we take the complex conjugate of $\mathcal{S}(p_0)$, interchange $\phi$ and $\varphi$, and replace $\ue^{\mathrm{i}\Omega j}$ with $\ue^{-\mathrm{i}\Omega j}$.
Then we can construct the propagator for $qB<0$, as follows:
\begin{equation}
\begin{split}
 \label{eq:SFd}
	\tilde S(x,x\prm)&=
		i\int\frac{\ud p_0\ud p_z}{(2\pi)^2}\sum_{l=-\infty}^\infty\sum_{k=1}^\infty
		\frac{1}{\p R^2N^2_{l,k}} \\
		&\quad \times\frac{\ue^{-\mathrm{i}(p_0+\Omega j -\mu)\D t+\mathrm{i}p_z\D z} }{p_0^2- \varepsilon^2+\mathrm{i}\epsilon}
		 \tilde{\mathcal S}(p_0),
\end{split}
\end{equation}
with
\begin{eqnarray}
 &\tilde{\mathcal S}(p_0)=
  \begin{pmatrix}
  		\tilde{\mathcal M}_+ & \tilde{\mathcal N} \\
   		-\tilde{\mathcal N} & \tilde{\mathcal M}_-
  \end{pmatrix},\\
 & \tilde{\mathcal M}_\pm = (\pm p_0+m)
 	\begin{pmatrix}
 		\vf^*\vf\prm & 0 \\
    	0 & \f^*\f\prm
 	\end{pmatrix},\\
  &  \tilde{\mathcal N } =
  	\begin{pmatrix}
  	 	-p_z\vf^*\vf\prm & -\mathrm{i}\sqrt{2|qB|\llk}\vf^*\f\prm \\
     	\mathrm{i}\sqrt{2|qB|\llk}\f^*\vf\prm & p_z\f^*\f\prm
  	\end{pmatrix}.
\end{eqnarray}
The sign difference in front of $\Omega j$ in $\mathcal{S}(p_0)$ and $\tilde{\mathcal{S}}(p_0)$ implies that $\Omega j$ can be regarded as an effective isospin chemical potential. %the Fermi surface of $u$-quarks ($d$-quarks) are pushed up (down).
The same is true for charged scalar particles, as seen in the dispersion~\eqref{eq:Epi}~\cite{Liu:2017spl}.
We should mention that this effective isospin chemical potential is induced only if both magnetic field and rotation are applied.
%Indeed, the $\mathcal{T}$-symmetry ($j\leftrightarrow -j$) is restored for $B=0$, and there is no effective chemical potential for $\Omega = 0$.
Indeed, when $B=0$, the symmetry $j\leftrightarrow -j$ is restored, so $\Omega j$ cannot be interpreted as an isospin chemical potential; 
and when $\Omega=0$, there is simply no such effective chemical potential. 

\section{Schwinger Phase}\label{sec:Sch}
The Dirac propagator in a constant magnetic field is in general gauge dependent (the results presented in Sec.~\ref{sec:Dirac} is for the symmetric gauge).
In Minkowski spacetime, the gauge dependence is completely incorporated by the Schwinger phase.
This is true also for the rotating frame as we show in the following, and the generalization to an arbitrary curved spacetime will be present elsewhere.

The Schwinger phase $\Theta(x,x')$ is defined through $S(x,x')=\ue^{\mathrm{i}\Theta(x,x')}S_{\rm inv}(x,x')$ so that $\Theta(x,x)=0$ is satisfied and $S_{\rm inv}$ is gauge invariant.
The condition $\Theta(x,x)=0$ means that the Schwinger phase is independent of the path connecting $x$ and $x'$.
As a biscalar, $\Theta(x,x')$ is invariant under coordinate transformation.
The expression in rotating frame can thus be obtained from the usual Schwinger phase in Minkowski spacetime.
For a constant $F_{\m\n}$ (i.e., $\nabla_\r F_{\m\n}=0$ with the covariant derivative $\nabla_\r$), the result is
\begin{equation}
\label{schwinger}
  \Theta(x,x')
  = - q\int_x^{x'}
  		\left[
  			A_\m(z)+\frac{1}{2}F_{\m\r}\nabla^\r_z s(z,x)
  		\right] \ud z^\m\,,
\end{equation}
where the integral is along an arbitrary path from $x$ to $x'$.
We have introduced the Synge's world function $s(a,b)$, which is half the squared geodesic distance between $a$ and $b$ \cite{Poisson:2011nh}:
\begin{equation}
 s(a,b)
 = \frac{1}{2}\int_0^1 g_{\m\n}(\xi(\t))\frac{\ud \xi^\m}{\ud \t}\frac{\ud \xi^\n}{\ud \t}\ud \t
\end{equation}
with $\xi^\m(0)=a^\m$ and $\xi^\m(1)=b^\m$.
This is reduced to $s(a,b)=(1/2)\eta_{\mu\nu}(a-b)^\mu(a-b)^\nu$ in Minkowski spacetime.
We can verify the path independence of $\F(x,x')$, by showing that the curl of the integrand in Eq.~\eqref{schwinger} vanishes after substituting the relation $\nabla^a_\m\nabla^a_\n s(a,b)=g_{\m\n}(a)$~\footnote{This relation holds for any spacetime with zero Riemann curvature.}.

When we choose the integral path as the geodesic from $x$ to $x'$, the second term in Eq.~(\ref{schwinger}) vanishes because $\nabla^\m_z s(z,x)$ is tangent to the geodesic.
We are left with a neat expression for the Schwinger phase
\begin{equation}
\label{schwinger2}
  \Theta(x,x')= -q\int_x^{x'} A_\m(z)\ud z^\m
\end{equation}
with geodesic being the integral path. See Appendix \ref{geodesicline} for an explicit representation of the geodesic in rotating frame.

To end this section, we show how the Schwinger phase can be factored out.
For simplicity, we focus on an unbounded system, where the fermion propagator is computed with the following replacement~\cite{Chen:2015hfc}:
\begin{equation}
\begin{split}
\label{eq:Landau}
 & \lambda\in \mathbb{N}\,,\quad
 \frac{1}{\pi R^2N^2_{lk}}\to\frac{|qB|}{2 \pi}\,,\quad -\lambda \leq l \leq -\lambda+N \
\end{split}
\end{equation}
with $N = \lfloor qBS/2 \pi\rfloor$.
We should note that $N$'s of $u$ and $d$ quarks are different because so are their charges: $q_u = 2e/3$ and $q_d = -e/3$.
Also we consider that magnetic field is strong enough to justify the LLL approximation.
Summing over all $l$'s under the replacement $N\to\infty$\,, then we arrive at%
\begin{equation}
\begin{split}
 \label{eq:LLLprop}
S_{\rm LLL}(x_1,x_2)
	& = \mathrm{i}\chi_\perp\int\frac{\ud p_0\ud p_z}{(2\pi)^2}
		\frac{\ue^{-\mathrm{i} (p_0-\Omega/2 -\mu)\D t+\mathrm{i}p_z\D z} }{p_0^2- p_z^2-m^2+\mathrm{i}\epsilon} \\
	& \quad \times(p_0\g^0-p_z\g^3+m)(1+\mathrm{i}\g^1\g^2)\,,
\end{split}
\end{equation}
%\begin{equation}
%\begin{split}
%\label{eq:chi}
%\chi_\perp(x_1,x_2)
%%	& =\sum_{l=0}^\infty\frac{1}{l!}\mathrm{e}^{il(\Delta \theta+\Omega\Delta t)} (\frac{1}{2}qBr_1r_2)^l\mathrm{e}^{-\frac{1}{4}qB(r_2^2+r_1^2)} \\
%%    &=\exp\left[\mathrm{e}^{i(\Delta \theta+\Omega\Delta t)}\frac{1}{2}qBr_1r_2-\frac{1}{4}qB(r_2^2+r_1^2)\right] \\
%%  &=\exp\left[-iq\int^{x_2}_{x_1}A_\mu\ud z^\mu-\frac{1}{4}qBC_\perp^2\right]\,.\\
%	& = \exp\biggl[
%			\frac{i}{2} qB r_1r_2\sin\Delta\vartheta
%	 		-\frac{1}{4}qB C^2_\perp
%	 		\biggr] \,, \\
% C^2_\perp
% &=r_2^2-2r_1r_2\cos\Delta \vartheta + r_1^2 \,,
%\end{split}
%\end{equation}
\begin{eqnarray}
\label{eq:chi}
& \displaystyle
 \chi_\perp(x_1,x_2)
 = \exp\biggl[
			\frac{\mathrm{i}}{2} qB r_1r_2\sin\Delta\vartheta
	 		-\frac{1}{4}qB C^2_\perp
	 	\biggr] \,, \\
\label{eq:Cperp}
& C^2_\perp
 =r_2^2-2r_1r_2\cos\Delta \vartheta + r_1^2 \,,
\end{eqnarray}
where we define $\Delta\vartheta = \Delta \theta+\Omega \Delta t$.
Here we notice that the first term in Eq.~\eqref{eq:chi} is the Schwinger phase with the integral path along the geodesic:
\begin{equation}
 -q\int^{x_2}_{x_1}A_\mu\ud z^\mu
  = \oh qB r_1r_2\sin\Delta \vartheta \,,
\end{equation}
where we employ the symmetric gauge.
It should be noticed here that the second term in Eq.~\eqref{eq:chi} is gauge invariant. 
We note that% and translational invariant.Indeed, the left-hand side is the geodesic distance from $x_1$ to $x_2$ is given by
\begin{equation}
  \begin{split}
%  \int^{x_2}_{x_1}\ud s
%  	=
  \int^1_0\sqrt{g_{\mu\nu} \frac{\ud x^\mu}{\ud\tau}\frac{\ud x^\nu}{\ud\tau}}\ud\tau
  	=\sqrt{\Delta t^2-C^2_\perp-\Delta z^2} \,.
  \end{split}
\end{equation}
This implies that $C_\perp $ is just the perpendicular distance measured in Minkowski spacetime. 
For $\O=0$, Eq.~\eqref{eq:LLLprop} reproduces the usual LLL Dirac propagator under a constant magnetic field in Minkowski spacetime.

\section {condensates}
In this section, we will use the two-flavor NJL model to analyze the chiral condensate and charged-pion condensate.
For zero current quark mass, the NJL Lagrangian in rotating frame is
\begin{equation}
\label{njllag}
\begin{split}
  \Lag_{\rm NJL}
  & =\bar\j i D \psi +\frac{G}{2}\Bigl[(\bar\j \j)^2+(\bar\j i\g^5\vec{\t}\j)^2\Bigr]\,, \\
  \mathrm{i} D
  & = \mathrm{i} \g^\m\nabla_\m +\mu_\B\gamma^0 \,.
\end{split}
\end{equation}
Here $\psi = (u,d)^T$ is the two flavor quark field, $\nabla_\m = \partial_\mu + \mathrm{i} Q A_\mu +\Gamma_\mu$ the covariant derivative, $\mu_\B$ (1/3 of) the baryon chemical potential, $Q=\diag(q_u,q_d) = \diag(2e/3,-e/3)$ the charge matrix in flavor space, and $\vec\tau$ the Pauli matrix.
%The four-fermion interaction in Lagrangian (\ref{njllag}) has the symmetry $U_\B(1)\otimes SU_{\rm V}(2)\otimes SU_{\rm A}(2)$.
In the mean field approximation, the one-loop effective action reads
\begin{equation}
\begin{split}
   \Gamma(\s,\vec\p)
   & =-\int\ud^4 x \frac{\sigma^2+\vec{\pi}^2}{2G}
   		-\mathrm{i}\,\Tr\ln(\mathrm{i}D-\sigma-\mathrm{i}\gamma^5\vec{\pi}\cdot\vec{\tau})\,,
 \end{split}
\end{equation}
where we introduce $\sigma=-G\langle \bar\psi\psi\rangle$ and $\vec \pi=-G\langle \bar\psi \mathrm{i}\gamma^5\vec\tau\psi\rangle$.
While the presence of the magnetic field spoils the $SU_{\rm V}(2)\otimes SU_{\rm A}(2)$ symmetry, its diagonal part remains.
This fact allows us to eliminate the neutral pion condensate by a chiral rotation. Hereafter we will always assume $\pi^0\equiv\pi^3=0$.

Now let us employ the imaginary-time formalism.
Then the effective action transforms into the thermodynamic potential
\begin{eqnarray}
	V_\eff(\s,\vec\p)&=&-\mathrm{i}\frac{1}{\beta V}\Gamma_{\rm E}(\s,\vec\p),
\end{eqnarray}
with $\G_E$ being the one-loop effective action after the Wick rotation to Euclidean spacetime.
For the current setup, it is practically hard to find the global minimum of $V_\eff(\s,\vec\pi)$, which corresponds to the ground state.
Instead, we will first calculate the minimum of $V_\eff(\s,\vec0)$, and then analyze its stability against the charged-pion fluctuation.
In other words, we perform the Ginzburg-Landau expansion up to the second order in the pion fields.
This is justified for the analysis around a second-order or a weak first-order phase transition, which is enough for our purpose.
The thermodynamic potential that we examine is thus written as
\begin{equation}
\label{eq:veff}
  \begin{split}
  V_\eff
  	& =V_\eff^{(0)}+V_\eff^{(2)}+\dots\,,\\
  V_\eff^{(0)}
    &=\frac{1}{\beta V}\int\ud^4 x_{\rm E} \frac{\sigma^2}{2G}
    -\frac{1}{\beta V}\Tr\ln(\mathrm{i}D-\sigma)\,,\\
  V_\eff^{(2)}
  	&=\frac{1}{\beta V}\int\ud^4 x_{\rm E} \frac{|\vec{\pi}|^2}{2G}
  		-\frac{1}{2\beta V}\Tr[(\mathrm{i}D-\sigma)^{-1}\gamma^5\vec{\pi}\cdot\vec{\tau}]^2\,.
  \end{split}
\end{equation}
The linear terms do not appear due to the invariance of $\G(\s,\vec\p)$ under the $\mathbb{Z}_2$ transformation $\vec\pi\ra-\vec\pi$.
We note that each term in Eq.~\eqref{eq:veff} is gauge invariant, as the gauge transformation leads to $A_\mu \to A_\mu + \partial_\mu\alpha(x)$ and $\p^{\pm}(x)\ra e^{\pm \mathrm{i}e\alpha(x)}\p^{\pm}(x)$.

Let us prepare several parameter choices in our NJL model analysis.
For the numerical calculation in this section, we choose the model parameters as
\begin{equation}
\label{eq:parameter1}
  G=11\Lambda^{-2},\ 17\Lambda^{-2}\,,\quad
  R=30\Lambda^{-1} \,.
\end{equation}
Besides, we note that the NJL model is nonrenormalizable.
Hereafter, when we carry out the momentum sum, we shall implicitly introduce the following smooth cutoff regulator~\cite{Gorbar:2011ya,Chen:2015hfc}:
\begin{equation}
\label{eq:reglu}
  f(k_f;\L,\d\L)
  =\frac{\sinh(\L/\d \L)}{\cosh(k_f/\d\L)+\cosh(\L/\d\L)}
\end{equation}
with $k_f = \sqrt{2q_fB\lambda + p_z^2}$ and $f=u,d$.
Here $\Lambda$ is the UV cutoff scale, and $\delta\Lambda$ characterizes the smoothness of the UV cutoff, which we choose
\begin{equation}
\label{eq:parameter2}
  \delta\Lambda=0.05\Lambda \,.
\end{equation}
The parameter choice in Eqs.~\eqref{eq:parameter1}-\eqref{eq:parameter2} is almost the same as that in Ref.~\cite{Chen:2017xrj}%
~\footnote{%
The former coupling $G=11\Lambda^{-2}$ in the two-flavor NJL model is the same as $G=22\Lambda^{-2}$ in the one-flavor case.%
}.
In the following, we focus on the zero density case $\mu_\B=0$ and the negatively finite density case $\mu_\B < 0$.
The reason why we adopt $\mu_\B<0$ will be explained in Section~\ref{sec:pion}.

\subsection{Chiral condensate}
\label{sec:chiralcond}

\begin{figure}
\centering
    \includegraphics[width=0.9\columnwidth]{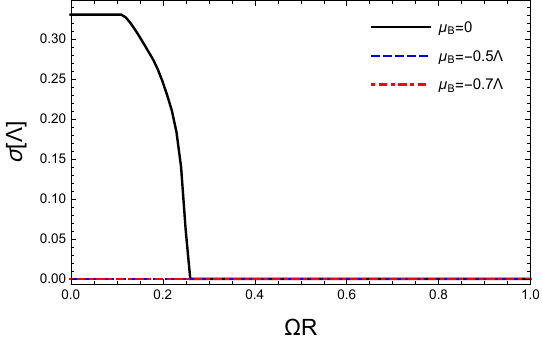}
  \vspace{1em}\\
    \includegraphics[width=0.9\columnwidth]{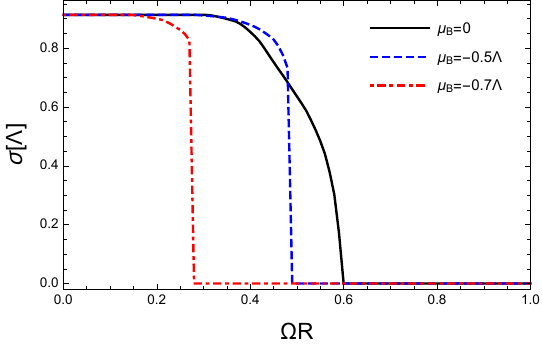}
    \caption{%
    Chiral condensate $\sigma$ as a function of the angular velocity for different baryon chemical potentials with $G=11\Lambda^{-2}$ (upper panel) and $G=17\Lambda^{-2}$ (lower panel).
    We employ $|q_d B| = 0.1\Lambda^2$ and the temperature is set to be zero.}
    \label{fig:chiral0}
\end{figure}

Due to the numerical difficulty to calculate the Ginzburg-Landau coefficients, we focus on an unbounded system, where Eq.~\eqref{eq:Landau} is applicable.
The calculation of the chiral condensate in a finite cylinder with no-flux boundary condition is represented in Appendix~\ref{app:cylinder}.
We also treat $\sigma$ to be independent of the spacetime coordinate.
Then the chiral condensate $\sigma$ is determined through the gap equation $\partial V^{(0)}_\eff/\partial \sigma = 0$, that is,
\begin{equation}
\begin{split}
\label{eq:gap}
  \frac{\sigma}{G}
    = \frac{\sigma}{S}\int \frac{\ud p}{2\pi}\sum_{\lambda, l, a, f} \frac{\alpha_\lambda}{\varepsilon}
    \biggl[
		\frac{1}{2}
		- n_F(\varepsilon_f - a\tilde{\mu}_f)
	\biggr]\,,
\end{split}
\end{equation}
%\begin{equation}
%\begin{split}
% \label{eq:V0}
% V_\eff^{(0)}
% & =\frac{\sigma^2}{2G}-\frac{1}{S}\int\frac{\ud p_z}{2\pi}
% 	\sum_{l,\l}\sum_{a=\pm}\sum_{f=u,d}(2-\delta_{0\lambda}) \\
% & \times
% 	\biggl[
% 		\frac{\varepsilon_f - \tilde\mu}{2}
% 		+\frac{1}{\beta}\ln\Bigl(1+\ue^{-\beta(\varepsilon_f-a\tilde\mu)}\Bigr)
% 	\biggr] \,,
%\end{split}
%\end{equation}
where we denote $n_F(x) = 1/(\ue^{\beta x} + 1)$, $\tilde\mu_f = \sgn(q_f)\Omega j+\mu_\B$, $\varepsilon_f = \sqrt{p_z^2 + 2q_f B\lambda + \sigma^2}$, and $\alpha_\lambda = 2-\delta_{0\lambda}$.
The factor $\alpha_\lambda$ comes from that $\Phi_{l+1}(-1,\tfrac{1}{2}|qB|r^2)=0$ for arbitrary $r$;
in the unbounded system either the spin-up or down cannot occupy the LLL.
We also note that the first and second terms in the brackets of Eq.~\eqref{eq:gap} correspond to the vacuum part and the effective finite-density part with chemical potential $\tilde\mu_f = \sgn(q_f)\Omega j+\mu_\B$, respectively.
At zero temperature, these terms (after the summation over $a$) are reduced as
\begin{equation}
\label{eq:zeroT}
 \sum_a \biggl[
		\frac{1}{2}
		- n_F(\varepsilon_f - a\tilde{\mu}_f)
	\biggr]
  \to
 \theta(\varepsilon_f-|\tilde{\mu}_f|) \,.
\end{equation}
For this reason, we expect that Eq.~\eqref{eq:gap} for large $\Omega$ or $\mu_\B$ has no nontrivial solution.
This reflects that the effective Fermi surface induced by $\tilde \mu$ suppresses the low-energy mode excitation, which is required to form the chiral condensate.
At the same time, we readily find that the scale to character the onset of the rotational effect is
\begin{equation}
 \mu_N = \Omega N \,,
\end{equation}
which is the rotation-induced effective chemical potential for the maximum angular momentum~\cite{Chen:2015hfc}.

In \fig{fig:chiral0}, the upper (lower) panel shows the numerical solution for Eq.~\eqref{eq:gap} at zero temperature with $G=11\Lambda^{-2}$ ($G=17\Lambda^{-2}$).
The result with $\mu_\B = 0$ and $G=11\Lambda^{-2}$ (the black line on the upper panel) is parallel to that obtained in Ref.~\cite{Chen:2015hfc}, expect for the number of flavors.
We confirm from Fig.~\ref{fig:chiral0} that the chiral condensate is more destructed either for larger $\Omega$ or larger $\mu_\B$, as we have explained below Eq.~\eqref{eq:zeroT}.
The destructions of the condensate by rotation correspond to the rotational magnetic inhibition~\cite{Chen:2015hfc}.

At the end of this subsection, we discuss the difference between the rotational effects on each flavor.
From Eq.~\eqref{eq:zeroT}, we find that a $j$-mode of $u$ quark receives the finite-density effect with $\tilde{\mu}_u = \Omega j -|\mu_\B|$, while the one of $d$ with $\tilde{\mu}_d = -\Omega j - |\mu_\B|$.
Hence, it might seem that the rotational effect on $u$ and $d$ behaves only as an effective isospin chemical.
On top of this, however, the rotational effect involves the aspect of the baryon chemical potential.
The main reason is because the upper bound of $l$ for each flavor are different, i.e., $l\lesssim N$ with $N \sim 2eBS/3$ for $u$ and with $N \sim eBS/3$ for $d$.
The rotational effect on $u$ thus differ from that on $\bar d$, which affect charged-pion condensate as we will discuss in the following.

\subsection{Charged-pion condensate}\label{sec:pion}

To calculate $V_\eff^{(2)}$, we make the following ansatz for the two-pion function:
\begin{equation}
\label{eq:ansatz}
 \pi^+(x')\pi^-(x)
 	=\exp\biggl[\mathrm{i}e\int^{x'}_{x} A_\mu\ud z^\mu\biggr]
 	  \tilde\pi^+(x')\tilde\pi^-(x) \,,
\end{equation}
where the integral path is along the geodesic between $x$ and $x'$, and $\tilde\pi^+(x')\tilde\pi^-(x)$ is gauge invariant.
In other words, the gauge dependent part of the correlator is assumed to be extracted only as the Wilson line. Note that although the geodesic is the most natural choice, other integral path is also possible, as in~\cite{Cao:2019ctl}, which results in different conclusions from ours. See the Appendix~\ref{app:another} for comparison of two different choices.
We emphasize that a similar ansatz is also employed to construct the gauge invariant two-point observables in the context of the usual electric superconductivity~\cite{Frohlich:1981yi,Bricmont:1983pq}. 
To proceed, we further suppose $\tilde\pi^+$ and $\tilde\pi^-$ to be constant spatially and temporally.
Then the second order term of the thermodynamic potential is written as
\begin{equation}
\begin{split}
\label{eq:Veff2}
  V_\eff^{(2)}=C^{(2)}\tilde\pi^+\tilde\pi^- \,,
\end{split}
\end{equation}
where $C^{(2)}$ is evaluated with the quark propagator.
It is important to note that $C^{(2)}$ is gauge independent since the Schwinger phases in the quark propagators compensate the Wilson line in Eq.~\eqref{eq:ansatz} exactly.

In vacuum, this $C^{(2)}$ could be regarded as an infrared energy squared of charged pions, as it is the coefficient in front of $\pi^+\pi^-$.
This interpretation is however not directly applied to the present system.
Indeed, although the charged pions has no baryon number, this $C^{(2)}$ are affected by the rotation-induced effective baryon chemical potential through the effect on the quark dynamics, as we argued below.

\begin{figure}
\centering
    \includegraphics[width=0.9\columnwidth]{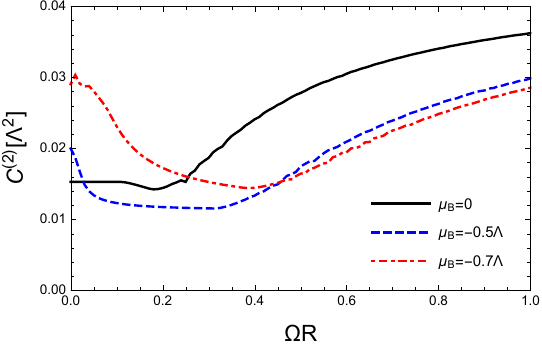}
    \vspace{1em}\\
    \includegraphics[width=0.9\columnwidth]{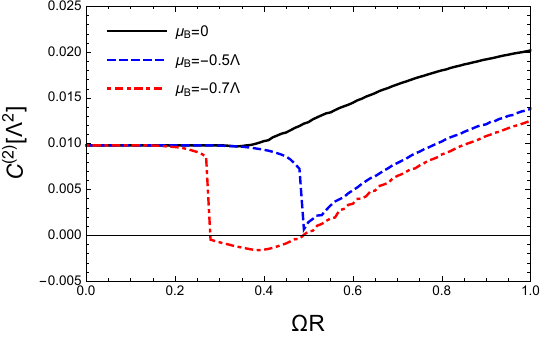}
    \caption{The Ginzburg-Landau coefficient $C^{(2)}$ as a function of the angular velocity with $G=11\Lambda^{-2}$ (upper panel) and $G=17\Lambda^{-2}$ (lower panel). We employ $|q_d B| = 0.1\Lambda^2$ and the temperature is set to be zero.}
    \label{fig:C2}
\end{figure}

In the Ginzburg-Landau analysis, $C^{(2)}<0$ characterises the instability of the $\s$ condensed state against the charged-pion fluctuation.
Thus this is the criterion for the phase transition towards the charged-pion condensation.
We can write down $C^{(2)}$, as follows:
\begin{eqnarray}
\label{eq:C2A1}
 &\displaystyle C^{(2)}
  = \frac{1}{2G} + \mathcal{C} \nonumber \\
 &\displaystyle \mathcal{C}
  = \frac{q_uB |q_d B|}{S}\int\frac{\ud p_z}{2\pi}
   	\sum_{\lambda_u,l_u} \sum_{\lambda_d,l_d} \sum_a
   	\frac{g_u-g_d}{\varepsilon_u\varepsilon_d(\varepsilon^2_u-\varepsilon^2_d)} \,,\ \
\end{eqnarray}
\begin{equation}
\begin{split}
\label{eq:gf}
 g_f
 &= \varepsilon_f
	\biggl[
		2q_f B\lambda_f I_1
		-\,\sgn\Bigl(j_u j_d\Bigr)
			\sqrt{2q_uB\lambda_u}\sqrt{2|q_dB|\lambda_d} I_2
	\biggr] \\
 &\quad \times
%		\tanh\frac{\beta (\xi-\Omega j-\mu_\B)}{2}\,,\\
		\biggl[
			\frac{1}{2}
			- n_F(\varepsilon_f - a\tilde{\mu}_f)
		\biggr] \,,\\
 I_1
 &= \int_0^\infty \ud r r \int_0^\infty\ud r' r'
	J_{j_u+j_d}\Bigl(\tfrac{1}{2}eBrr'\Bigr) \\
 \times &
	\biggl[
		\Phi^u_\uparrow(r) \Phi^d_\downarrow(r)
		\Phi^u_\uparrow(r') \Phi^d_\downarrow(r')
  		+\Phi^u_\downarrow(r) \Phi^d_\uparrow(r)
  		 \Phi^u_\downarrow(r') \Phi^d_\uparrow(r')
  	\biggr] \,,\\
 I_2
 &=
	2\int_0^\infty \ud r r \int_0^\infty\ud r' r'
		J_{j_u+j_d}\Bigl(\tfrac{1}{2}eBrr'\Bigr)  \\
 &\quad \times
		\Phi^u_\uparrow(r) \Phi^d_\downarrow(r)
		\Phi^u_\downarrow(r')\Phi^d_\uparrow(r') \,.
\end{split}
\end{equation}
Here we denote $j_f = l_f + 1/2$ as the total angular momentum, and $J_l(x)$ represents the first kind of the Bessel function.
Also we introduced shorthand notations: $\Phi^f_\uparrow(r) = \Phi_{l_f}(\lambda_f,\tfrac{1}{2} |q_f B| r^2)$ and $\Phi^f_\downarrow(r) = \Phi_{l_f+1}(\lambda_f-1,\tfrac{1}{2} |q_f B| r^2)$.
%\begin{equation}\label{eq:C2A1}
% C^{(2)}
% = \frac{1}{2G}
%   +\frac{q_uB |q_d B|}{S}\int\frac{\ud p_z}{2\pi}
%   	\sum_{\lambda \lambda\,'l l'}\sum_{a,b=\pm}
%   	G(a\varepsilon_u,b\varepsilon_d)
%\end{equation}
%with
%\begin{widetext}
%\begin{eqnarray}
%G(\xi,\zeta)&=&\frac{1}{2 \xi \zeta}\frac{g(\xi)-g(\zeta)}{\xi-\zeta}\,,\\
%g(\xi)&=&
%	\Bigl[
%		(p_z^2+\sigma^2-\xi^2)I_1
%		-\,\sgn(jj')\sqrt{2q_uB\lambda}\sqrt{2|q_dB|\lambda'} I_2
%	\Bigr]
%%		\tanh\frac{\beta (\xi-\Omega j-\mu_\B)}{2}\,,\\
%	\biggl(
%		\frac{1}{2} - \frac{1}{\ue^{\beta (\xi-\Omega j-\mu_\B)} + 1}
%	\biggr)\,,\\
%I_1&=& \int_0^\infty \ud r r \int_0^\infty\ud r' r'
%	J_{l+l'+1}\Bigl(\tfrac{1}{2}eBrr'\Bigr)
%	\biggl[
%		\F_{l}^u(\lambda,r) {\F}^d_{l'+1}({\l}'-1,r) \F_{l}^{u}(\lambda,r'){\F}^{d}_{l'+1}({\l}'-1,r') \nonumber\\
%  &&
%	\qquad\qquad\qquad\qquad\qquad\qquad\qquad\qquad\qquad
%  		+\F^u_{l+1}(\l-1,r) {\F}^d_{l'}({\l}',r) \F_{l+1}^{u}(\l-1,r') {\F}^{d}_{l'}(\l',r')
%  	\biggr] \,,\\
%I_2&=&
%	2\int_0^\infty \ud r r \int_0^\infty\ud r' r'
%		J_{l+l'+1}\Bigl(\tfrac{1}{2}eBrr'\Bigr)
%		\F_{l}^u(\lambda,r){\F}^d_{l'+1}({\l}'-1,r) \F_{l+1}^{u}(\l-1,r'){\F}^{d}_{l'}(\l',r') \,.
%\end{eqnarray}
%\end{widetext}
We note that again the (effective) finite-density effect enters into Eq.~\eqref{eq:C2A1}, as well as the gap equation~\eqref{eq:gap}.
In this sense, similarly to the chiral condensate $\sigma$, the low-energy mode accumulation is necessary to generate the pion condensate;
otherwise, we have $C_2\simeq (2G)^{-1} > 0$, and thus no pion condensate is generated.
In Fig.~\ref{fig:C2}, we present the zero-temperature numerical results of $C^{(2)}$ with $|q_d B| = 0.1\Lambda^{-2}$.
The results for $G=11\Lambda^{-2}$ ($G=17\Lambda^{-2}$) are plotted in the upper (lower) panel.
We note that there is a numerical cost to increase the precisions of $I_1$ and $I_2$, which are the multi-integration of products of highly oscillating functions.
The nonsmooth behaviors in Fig.~\ref{fig:C2} is hence the numerical artifacts that comes from less precisions of such evaluations.

Let us first look at the $\mu_\B=0$ case, which corresponds to the black lines in Fig.~\ref{fig:C2}.
Although the strong magnetic field $eB=0.1\Lambda^2$ generates a lot of the zero modes (the Landau quantization), it also increases a constituent quark mass $\sigma$ (the magnetic catalysis).
That is, the low-energy modes never contribute to $C^{(2)}$ unless the chiral restoration takes place due to an effective finite density effect (the rotational magnetic inhibition).
This is nothing but a Silver Blaze phenomenon;
the rotational effect on microscopic quantities at zero temperature becomes visible only if the effective chemical potential $\mu_\eff \sim \O N$ exceeds the infrared threshold $\sigma(\Omega=0)$.
This is the reason why the $\mu_\B=0$ case in Fig.~\ref{fig:C2} involves plateau behaviors around small $\Omega$ regions.
Also we have numerically checked that at zero baryon chemical potential, arbitrary strong $B$ leads to $C^{(2)}>0$ irrelevantly to the strength of $\Omega$.
At the zero density system, hence, the charged-pion condensate is disfavored~\cite{Cao:2015xja,Adhikari:2018fwm}.
This fact is consistent with the argument based on the spin-alignment of $u$ and $\bar d$ in magnetic field and rotation;
two fermions with the parallely aligned spins cannot form the spin-0 composite states.

On the other hand, for a negatively large $\mu_\B$, the chiral condensate is washed out, as seen in Fig.~\ref{fig:chiral0}.
In this case, the low-energy modes can contribute to $C^{(2)}$.
Therefore $C^{(2)}$ decreases only if $\sigma$ also does, which we find from Figs.~\ref{fig:chiral0} and~\ref{fig:C2}.
This is plausible if we would regard $C^{(2)}$ as the charged-pion mass, which is partially determined by the chiral condensate~\cite{GellMann:1968rz} except for the magnetic-induced part.
We also mention that the plateaus in the lower panel represent a Silver Blaze phenomena for the effective chemical potential $\mu_N + \mu_\B = \Omega N + \mu_\B$.

Furthermore, when the coupling constant is strong enough, $|\mathcal{C}|$ exceeds $(2G)^{-1}$ and thus the pion condensation can be realized.
In Fig.~\ref{fig:C2}, the relevant case is only the result with $\mu_\B = -0.7\Lambda^2$ and $G=17\Lambda^{-2}$ (the red line in the lower panel).
This is understandable in the sense that condensate cannot be formed in a weakly interacting system and mimics a BCS to BEC type crossover. 
We note that a large $\mu_\B$ that overwhelms the rotational effects would always destroy mesonic condensations.
Only if we choose a negative $\mu_\B$ to let $\sigma$ disappear earlier, there will be a window allowing charged-pion condensation.

Here we explain why the negative chemical potential is required for the charged-pion condensation.
As we have mentioned already in Section~\ref{sec:chiralcond}, rotation affects each flavor not only as an effective isospin chemical potential but also as an effective baryon chemical potential.
Let us schematically define them as $\mu^\text{rot}_u = \mu^\text{rot}_\text{isospin} + \mu^\text{rot}_\text{baryon}$ and $\mu^\text{rot}_d = -\mu^\text{rot}_\text{isospin} + \mu^\text{rot}_\text{baryon}$.
For $\mu_\B = 0$, there is a mismatch of the Fermi surfaces of $u$ and $\bar{d}$ (namely, $\mu^\text{rot}_u-|\mu^\text{rot}_d|=2\mu^\text{rot}_\text{baryon}$), and the charged-pion condensate $\langle\pi^+\rangle = \langle \bar d \gamma^5 u \rangle$ would have a finite momentum~\cite{Yamamoto:2014lia}.
Such a condensate is inconsistent with our homogeneous ansatz as Eq.~\eqref{eq:Veff2}.
This is the reason why we get $C^{(2)}>0$ for $\mu_\B=0$.
However, if we apply $\mu_\B<0$ to compensate $\mu^\text{rot}_\text{baryon}>0$, the Fermi-surface mismatch is eliminated. 
Then, the rotating magnetized system effectively behaves as a purely isospin density system, where a homogeneous charged-pion condensate forms~\cite{Son:2000xc,He:2005nk,Sun:2007fc,He:2006tn}.
Indeed, it can be checked that the critical baryon chemical potential $|\mu_\B| \sim 0.7\Lambda$ is of the same order as $\mu_N = \Omega N$~\cite{Chen:2015hfc}.
The above argument is based on the dynamics of quarks inside pion.
This is the stark difference from Ref.~\cite{Liu:2017spl}, which focuses on the pion dynamics.

\section{Discussion}
In this paper, we revisited the possibility of a charged-pion condensation in a magnetic field with rotation.
We first discussed the gauge dependence of the Dirac propagator in curved spacetime and show that the Schwinger phase is given by a Wilson line along the geodesic.
Then we calculated the second-order coefficient in charged-pion field in the Ginzburg-Landau expansion of the two flavor NJL model.
We find that, at zero baryon chemical potential, the charged-pion condensation is unlikely to happen in a wide region of the parameters.
The essential reason is because the interplay between magnetic field and rotation generates, in addition to an effective isospin chemical potential, an effective baryon chemical potential, which implies a mismatch of the Fermi surfaces of $u$ and $\bar{d}$.
However, turning on a negative baryon chemical potential to compensate such a mismatch, we do find that charged-pion condensation may happen for certain values of angular velocity and magnetic field.
Our result reveals that the microscopic structure of pion is important due to the presence of an effective baryon chemical potential, which is not introduced in the pion dynamics.

We should mention that our analysis does not totally exclude the possibility of charged-pion condensation.
We mainly focus on the homogeneous charged-pion condensate, and show that such condensate is formed only in a system with a large chemical potential. 
Under large rotation, the system becomes highly inhomogeneous.
This could for example yield the situation that a vortex structure with the condensate could locate only at the region very close to boundary~\cite{Guo:2021gbz}.
Since it is quite nontrivial to incorporate such inhomogeneous configuration in our Ginzburg-Landau analysis, we will leave it for future study.

As a by-product, we also studied the chiral condensate in a finite cylinder with no-flux boundary condition in Appendix~\ref{app:cylinder}, as an extension of the analysis in Ref.~\cite{Chen:2017xrj} to a rotating system.
We find that due to the centrifugal force, the chiral condensate is suppressed near the boundary, and there is a nontrivial competition between the rotational magnetic inhibition and the surface magnetic catalysis. 

Some comments are in order:
(1)
Our Ginzburg-Landau analysis would not be rigorous for a large $|\mu_\B|$, which could leads to a strong first-order phase transition.
Nevertheless, our numerical studies provide a clearer underlying picture of the charged-pion condensation in finite $eB$ and $\Omega$.
(2) The charged-pion condensate triggers also an electric superconductivity, which may repel the magnetic field.
This disfavors the formation of the charged-pion condensate itself, especially in small magnetic field.
For large magnetic field, a vortex lattice may form to accommodate the magnetic field, if it is a type-II superconductor.
It will be interesting to explore these phenomena.
(3) Our analysis suggests that the condensation of charged spin-one particles (such as $\rho$) may be favored by a rotation with a magnetic field.
This mechanism differs from the ones in pure magnetic field and in finite isospin chemical potential;
The former may suffer from the constraint due to Vafa-Witten theorem~\cite{Hidaka:2012mz};
In the latter case, the inevitable pion condensate may make $\rho$ harder to condense~\cite{Brauner:2016lkh}.
We emphasize that the $\rho$ condensation under rotation is not prohibited by the Vafa-Witten theorem, because the presence of rotation violates the positivity of the Dirac determinant.
We leave this topic to future works.

\section*{Acknowledgment}
We thank Gaoqing~Cao, Kenji~Fukushima, Lianyi~He, and Dirk-Hermann~Rischke for useful discussions.
This work is supported by the National Key Research and Development Program of China (Grant No. 2022YFA1604900), the  Natural Science Foundation of China (Grant No.12247133,  No. 12225502 and No. 12075061), and the Natural Science Foundation of Shanghai (Grant No. 20ZR1404100).

\appendix

\section{Geodesic curve in a rotating frame}\label{geodesicline}
We derive the geodesic between two points $x_1$ and $x_2$ in a rotating frame in this appendix. The geodesic equation is
\begin{eqnarray}
\frac{\ud^2x^\mu}{\ud \tau^2}+\Gamma^\mu_{\nu\rho}\frac{\ud x^\nu}{\ud \tau}\frac{\ud x^\rho}{\ud \tau}=0,
\end{eqnarray}
where $\s$ is the affine parameter. For the $t$ and $z$ coordinates, the solutions are
\begin{eqnarray}
t(\tau)=t_1+(t_2-t_1)\tau,\nonumber\\
z(\tau)=z_1+(z_2-z_1)\tau,
\end{eqnarray}
which is the same as that in the Minkowski coordinate. For the transverse coordinates,
\begin{equation}
\begin{split}
\frac{\ud^2 x}{\ud \tau^2}-\Omega^2\Delta t^2 x-2\Omega\Delta t \frac{\ud y}{\ud \tau}&=0,\\
\frac{\ud^2 y}{\ud \tau^2}-\Omega^2\Delta t^2 y-2\Omega\Delta t \frac{\ud x}{\ud \tau}&=0,
\end{split}
\end{equation}
where $\Delta t=t_2-t_1$. Introducing $\xi=x+iy$ we get
\begin{eqnarray}
\ddot\xi-\Omega^2\Delta t^2 \xi+2i\Omega\Delta t \dot \xi=0.
\end{eqnarray}
The solution is
\begin{eqnarray}
\xi=C_1\ue^{-i(\Omega\Delta t \tau+\phi_1)}+C_2\ue^{-i(\Omega\Delta t \tau+\phi_2)}\tau,
\end{eqnarray}
i.e.,
\begin{equation}
\begin{split}
x(\tau)&=C_1\cos(\Omega\Delta t\tau+\phi_1)+C_2\tau \cos(\Omega\Delta t\tau+\phi_2),\\
y(\tau)&=-C_1\sin(\Omega\Delta t\tau+\phi_1)-C_2\tau \sin(\Omega\Delta t\tau+\phi_2),
\end{split}
\end{equation}
where $C_1$, $C_2$, $\phi_1$ and $\phi_2$ can be determined by the condition
\begin{equation}
\begin{split}
x(0)&=x_1, \quad x(1)=x_2,\\
y(0)&=y_1, \quad y(1)=y_2.
\end{split}
\end{equation}
Finally, we have
\begin{equation}
  \begin{split}
  C_1&=r_1,\;\;\;  \phi_1=-\theta_1,\nonumber\\
    C_2^2&=r_2^2-2r_1r_2\cos(\Delta \theta+\Omega\Delta t)+r_1^2\equiv C_\perp^2,\nonumber\\
  \phi_2&=-\O\D t-\arctan\frac{r_1\sin(\theta_1-\Omega\Delta t)-r_2\sin\theta_2}{r_1\cos(\theta_1-\Omega\Delta t)-r_2\cos\theta_2}.\nonumber
  \end{split}
\end{equation}

\section{Inhomogenous Chiral Condensate in a Cylinder}\label{app:cylinder}

\begin{figure}
    \includegraphics[width=0.9\columnwidth]{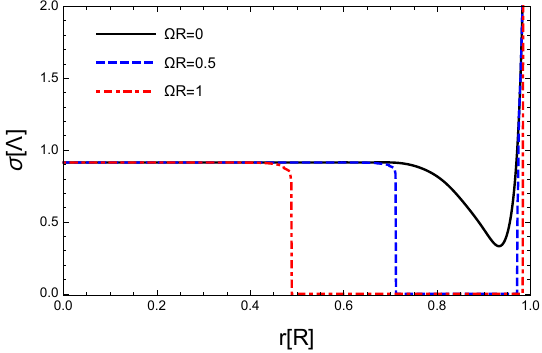}
    \caption{Chiral condensate in a finite cylinder as a function of the radial coordinate $r$ for the choice of $R = 30\Lambda^{-1}$ and $|q_dB|=0.1\Lambda^2$. The temperature is zero.}
    \label{fig:chiral}
\end{figure}
In a finite cylinder with transverse radius $R$, the thermodynamic potential $V_\eff^{(0)}$ becomes the functional of $\sigma(r)$ due to the breaking of the translational invariance.
Accordingly, the gap equation for $\s(r)$ reads $\delta V_\eff^{(0)}/\delta \sigma(r) = 0$.
Now we employ the local density approximation, where $|\partial_r \sigma(r)|$ is supposed to be negligible compared with $\sigma(r)^2$~\cite{Jiang:2016wvv}.
Then the gap equation is reduced to
\begin{equation}
\begin{split}
 \frac{\sigma(r)}{G}
 & =\int\frac{\ud p_z}{2\pi}\sum_{l,k}\sum_{a=}\sum_{f=u,d}
 	\frac{\sigma(r)}{2\varepsilon_f(r)}
\\
   & \quad \times
 	\biggl[
		\frac{1}{2}
		- n_F(\varepsilon_f(r) - a\tilde{\mu}_f)
	\biggr]
 	\frac{[\Phi^f_\uparrow(r)]^2
 		  +[\Phi^f_\downarrow(r)]^2}{\pi R^2N_{lk}^2}
 \,,
\end{split}
\end{equation}
where $\varepsilon_f(r) =\sqrt{p_z^2 + 2|q_f B|\lambda_{l,k} + \sigma(r)^2}$.
Note that $\llk$ depends on $q$ when a boundary condition is imposed.
Here we adopt the same boundary condition as~\cite{Chen:2017xrj}
\begin{equation}
\begin{split}
 &\Phi^f_\uparrow(R) = 0 \quad\text{for}\quad l\geq 0 \\
 &\Phi^f_\downarrow(R) = 0 \quad\text{for}\quad l\leq -1 \,.
\end{split}
\end{equation}
This is one of the conditions so that the incoming $u$- and $d$-quark current fluxes (i.e., the isospin current flux) vanishes individually at the boundary $r=R$, or equivalently, so that the hermiticity of the Hamiltonian is respected.
%The UV regulator is chosen as
%\begin{equation}
%\label{reglu2}
%  f_q(|\bp|;\L,\d\L)=\frac{\sinh(\L/\d \L)}{\cosh(|\bp|/\d\L)+\cosh(\L/\d\L)}
%\end{equation}
%with $|\bp|=\sqrt{2qB\lambda_{l,k}+p_z^2}$.
In the numerical calculation we choose the regulator~\eqref{eq:reglu} and the parameters~\eqref{eq:parameter1},~\eqref{eq:parameter2}%
~\footnote{
In this section, we employ only the larger coupling constant $G=17\Lambda^{-2}$.%
}.

The result for $\s(r)$ is shown in Fig.~\ref{fig:chiral}.
We also show the case with $\O=0$, which has been thoroughly studied in Ref.~\cite{Chen:2017xrj}.
The surface magnetic catalysis is clearly seen%
~\footnote{%
The numerical result at the vicinity of the boundary obviously contradicts the local density approximation.
This strong enhancement of $\sigma$ however comes from a number of accumulating modes at $r\sim R$.
In this sense, this magnetic response is expected to be qualitatively unchanged even with stronger inhomogeneity of $\sigma$ (see Ref.~\cite{Chen:2017xrj} for a detailed discussion).%
}.
Once the rotation is turned on, the rotational suppression first occurs in the region far away from the rotating axis where the centrifugal force is large.
Also the rotational effect is enhanced toward the center, as $\O$ increases.
On the other hand, at the vicinity of the boundary, the chiral condensate is amplified by the surface magnetic catalysis, but with its effective region shrinking with increasing $\O$; thus we observe a coexistence of and competition between the rotational magnetic inhibition and surface magnetic catalysis~\cite{Chen:2015hfc,Chen:2017xrj}.

\begin{figure}
\centering
    \includegraphics[width=0.9\columnwidth]{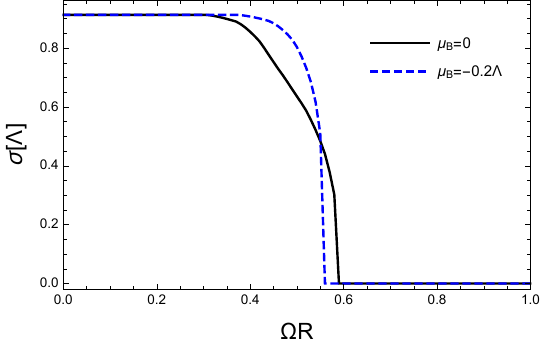}
    \vspace{1em} \\
    \includegraphics[width=0.93\columnwidth]{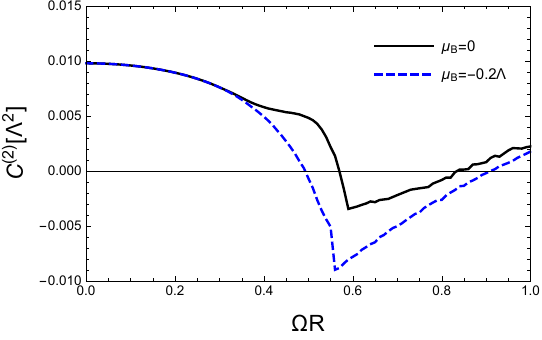}
    \caption{Chiral condensate $\sigma$ and Ginzburg-Landau coefficient $C^{(2)}$ as a function of the angular velocity for $|q_dB|=0.1\Lambda^2$ with the integral path given in Eq.~\eqref{eq:path2}.
    The temperature is zero, and we choose $G=17\Lambda^{-2}$.}
    \label{fig:A2}
\end{figure}

\section{Another Choice of the Integral Path}\label{app:another}

In Section.~\ref{sec:Sch} we have shown that the integral path in the Schwinger phase should be the geodesic.
For the sake of comparison, we in this appendix choose another integral path for the Schwinger phase. This also results in a corresponding change of the integral path in Eq.~\eqref{eq:ansatz}. This path from $x_1$ to $x_2$ is specified by
\begin{equation}\label{eq:path2}
\begin{split}
   (t_1,x_1,y_1,z_1)&\to(t_1,0,0,z_1)\to(t_2,0,0,z_2)\\
   &\to(t_2,x_1,y_1,z_2)\to(t_2,x_2,y_2,z_2),
   \end{split}
\end{equation}
where the arrows mean being connected by geodesics.
Then the integral is
\begin{equation}
 \mathrm{i}q\int^{x_2}_{x_1}A_\mu\ud z^\mu=-\frac{\mathrm{i}}{2} qB r_1r_2\sin\Delta \theta,
\end{equation}
which has the same expression as the Schwinger phase in the Minkowski coordinate, except that the coordinates here correspond to rotating frame. This formula is used in Ref.~\cite{Cao:2019ctl}.

The numerical result in this case is shown in Fig.~\ref{fig:A2}.
We find that $C^{(2)}$ begins to decrease with increasing $\Omega$, and the charged-pion condensation can happen without adding a negative baryon chemical potential.
However, we still find that in certain region a negative baryon chemical potential can catalyze the charged-pion condensation.

%%%%%%%%%%%%%%%%%%%%%%%%%%%%%%%%%%%%%%%%%%%%%%%%%%%%%%%%%%%%%%%%%%%
\bibliography{ref}

\begin{thebibliography}{60}%
\makeatletter
\providecommand \@ifxundefined [1]{%
 \@ifx{#1\undefined}
}%
\providecommand \@ifnum [1]{%
 \ifnum #1\expandafter \@firstoftwo
 \else \expandafter \@secondoftwo
 \fi
}%
\providecommand \@ifx [1]{%
 \ifx #1\expandafter \@firstoftwo
 \else \expandafter \@secondoftwo
 \fi
}%
\providecommand \natexlab [1]{#1}%
\providecommand \enquote  [1]{``#1''}%
\providecommand \bibnamefont  [1]{#1}%
\providecommand \bibfnamefont [1]{#1}%
\providecommand \citenamefont [1]{#1}%
\providecommand \href@noop [0]{\@secondoftwo}%
\providecommand \href [0]{\begingroup \@sanitize@url \@href}%
\providecommand \@href[1]{\@@startlink{#1}\@@href}%
\providecommand \@@href[1]{\endgroup#1\@@endlink}%
\providecommand \@sanitize@url [0]{\catcode `\\12\catcode `\$12\catcode
  `\&12\catcode `\#12\catcode `\^12\catcode `\_12\catcode `\%12\relax}%
\providecommand \@@startlink[1]{}%
\providecommand \@@endlink[0]{}%
\providecommand \url  [0]{\begingroup\@sanitize@url \@url }%
\providecommand \@url [1]{\endgroup\@href {#1}{\urlprefix }}%
\providecommand \urlprefix  [0]{URL }%
\providecommand \Eprint [0]{\href }%
\providecommand \doibase [0]{https://doi.org/}%
\providecommand \selectlanguage [0]{\@gobble}%
\providecommand \bibinfo  [0]{\@secondoftwo}%
\providecommand \bibfield  [0]{\@secondoftwo}%
\providecommand \translation [1]{[#1]}%
\providecommand \BibitemOpen [0]{}%
\providecommand \bibitemStop [0]{}%
\providecommand \bibitemNoStop [0]{.\EOS\space}%
\providecommand \EOS [0]{\spacefactor3000\relax}%
\providecommand \BibitemShut  [1]{\csname bibitem#1\endcsname}%
\let\auto@bib@innerbib\@empty
%</preamble>
\bibitem [{\citenamefont {Kharzeev}\ \emph {et~al.}(2008)\citenamefont
  {Kharzeev}, \citenamefont {McLerran},\ and\ \citenamefont
  {Warringa}}]{Kharzeev:2007jp}%
  \BibitemOpen
  \bibfield  {author} {\bibinfo {author} {\bibfnamefont {D.~E.}\ \bibnamefont
  {Kharzeev}}, \bibinfo {author} {\bibfnamefont {L.~D.}\ \bibnamefont
  {McLerran}},\ and\ \bibinfo {author} {\bibfnamefont {H.~J.}\ \bibnamefont
  {Warringa}},\ }\bibfield  {title} {\bibinfo {title} {{The Effects of
  topological charge change in heavy ion collisions: `Event by event P and CP
  violation'}},\ }\href {https://doi.org/10.1016/j.nuclphysa.2008.02.298}
  {\bibfield  {journal} {\bibinfo  {journal} {Nucl. Phys.}\ }\textbf {\bibinfo
  {volume} {A803}},\ \bibinfo {pages} {227} (\bibinfo {year} {2008})},\ \Eprint
  {https://arxiv.org/abs/0711.0950} {arXiv:0711.0950 [hep-ph]} \BibitemShut
  {NoStop}%
%%CITATION = ARXIV:0711.0950;%%
\bibitem [{\citenamefont {Fukushima}\ \emph {et~al.}(2008)\citenamefont
  {Fukushima}, \citenamefont {Kharzeev},\ and\ \citenamefont
  {Warringa}}]{Fukushima:2008xe}%
  \BibitemOpen
  \bibfield  {author} {\bibinfo {author} {\bibfnamefont {K.}~\bibnamefont
  {Fukushima}}, \bibinfo {author} {\bibfnamefont {D.~E.}\ \bibnamefont
  {Kharzeev}},\ and\ \bibinfo {author} {\bibfnamefont {H.~J.}\ \bibnamefont
  {Warringa}},\ }\bibfield  {title} {\bibinfo {title} {{The Chiral Magnetic
  Effect}},\ }\href {https://doi.org/10.1103/PhysRevD.78.074033} {\bibfield
  {journal} {\bibinfo  {journal} {Phys. Rev.}\ }\textbf {\bibinfo {volume}
  {D78}},\ \bibinfo {pages} {074033} (\bibinfo {year} {2008})},\ \Eprint
  {https://arxiv.org/abs/0808.3382} {arXiv:0808.3382 [hep-ph]} \BibitemShut
  {NoStop}%
%%CITATION = ARXIV:0808.3382;%%
\bibitem [{\citenamefont {Huang}(2016)}]{Huang:2015oca}%
  \BibitemOpen
  \bibfield  {author} {\bibinfo {author} {\bibfnamefont {X.-G.}\ \bibnamefont
  {Huang}},\ }\bibfield  {title} {\bibinfo {title} {{Electromagnetic fields and
  anomalous transports in heavy-ion collisions --- A pedagogical review}},\
  }\href {https://doi.org/10.1088/0034-4885/79/7/076302} {\bibfield  {journal}
  {\bibinfo  {journal} {Rept. Prog. Phys.}\ }\textbf {\bibinfo {volume} {79}},\
  \bibinfo {pages} {076302} (\bibinfo {year} {2016})},\ \Eprint
  {https://arxiv.org/abs/1509.04073} {arXiv:1509.04073 [nucl-th]} \BibitemShut
  {NoStop}%
%%CITATION = ARXIV:1509.04073;%%
\bibitem [{\citenamefont {Hattori}\ and\ \citenamefont
  {Huang}(2017)}]{Hattori:2016emy}%
  \BibitemOpen
  \bibfield  {author} {\bibinfo {author} {\bibfnamefont {K.}~\bibnamefont
  {Hattori}}\ and\ \bibinfo {author} {\bibfnamefont {X.-G.}\ \bibnamefont
  {Huang}},\ }\bibfield  {title} {\bibinfo {title} {{Novel quantum phenomena
  induced by strong magnetic fields in heavy-ion collisions}},\ }\href
  {https://doi.org/10.1007/s41365-016-0178-3} {\bibfield  {journal} {\bibinfo
  {journal} {Nucl. Sci. Tech.}\ }\textbf {\bibinfo {volume} {28}},\ \bibinfo
  {pages} {26} (\bibinfo {year} {2017})},\ \Eprint
  {https://arxiv.org/abs/1609.00747} {arXiv:1609.00747 [nucl-th]} \BibitemShut
  {NoStop}%
%%CITATION = ARXIV:1609.00747;%%
\bibitem [{\citenamefont {Kharzeev}\ \emph {et~al.}(2016)\citenamefont
  {Kharzeev}, \citenamefont {Liao}, \citenamefont {Voloshin},\ and\
  \citenamefont {Wang}}]{Kharzeev:2015znc}%
  \BibitemOpen
  \bibfield  {author} {\bibinfo {author} {\bibfnamefont {D.~E.}\ \bibnamefont
  {Kharzeev}}, \bibinfo {author} {\bibfnamefont {J.}~\bibnamefont {Liao}},
  \bibinfo {author} {\bibfnamefont {S.~A.}\ \bibnamefont {Voloshin}},\ and\
  \bibinfo {author} {\bibfnamefont {G.}~\bibnamefont {Wang}},\ }\bibfield
  {title} {\bibinfo {title} {{Chiral magnetic and vortical effects in
  high-energy nuclear collisions: A status report}},\ }\href
  {https://doi.org/10.1016/j.ppnp.2016.01.001} {\bibfield  {journal} {\bibinfo
  {journal} {Prog. Part. Nucl. Phys.}\ }\textbf {\bibinfo {volume} {88}},\
  \bibinfo {pages} {1} (\bibinfo {year} {2016})},\ \Eprint
  {https://arxiv.org/abs/1511.04050} {arXiv:1511.04050 [hep-ph]} \BibitemShut
  {NoStop}%
%%CITATION = ARXIV:1511.04050;%%
\bibitem [{\citenamefont {Gusynin}\ \emph {et~al.}(1994)\citenamefont
  {Gusynin}, \citenamefont {Miransky},\ and\ \citenamefont
  {Shovkovy}}]{Gusynin:1994re}%
  \BibitemOpen
  \bibfield  {author} {\bibinfo {author} {\bibfnamefont {V.~P.}\ \bibnamefont
  {Gusynin}}, \bibinfo {author} {\bibfnamefont {V.~A.}\ \bibnamefont
  {Miransky}},\ and\ \bibinfo {author} {\bibfnamefont {I.~A.}\ \bibnamefont
  {Shovkovy}},\ }\bibfield  {title} {\bibinfo {title} {{Catalysis of dynamical
  flavor symmetry breaking by a magnetic field in (2+1)-dimensions}},\ }\href
  {https://doi.org/10.1103/PhysRevLett.76.1005, 10.1103/PhysRevLett.73.3499}
  {\bibfield  {journal} {\bibinfo  {journal} {Phys. Rev. Lett.}\ }\textbf
  {\bibinfo {volume} {73}},\ \bibinfo {pages} {3499} (\bibinfo {year}
  {1994})},\ \bibinfo {note} {[Erratum: Phys. Rev. Lett.76,1005(1996)]},\
  \Eprint {https://arxiv.org/abs/hep-ph/9405262} {arXiv:hep-ph/9405262
  [hep-ph]} \BibitemShut {NoStop}%
%%CITATION = HEP-PH/9405262;%%
\bibitem [{\citenamefont {Gusynin}\ \emph {et~al.}(1996)\citenamefont
  {Gusynin}, \citenamefont {Miransky},\ and\ \citenamefont
  {Shovkovy}}]{Gusynin:1995nb}%
  \BibitemOpen
  \bibfield  {author} {\bibinfo {author} {\bibfnamefont {V.~P.}\ \bibnamefont
  {Gusynin}}, \bibinfo {author} {\bibfnamefont {V.~A.}\ \bibnamefont
  {Miransky}},\ and\ \bibinfo {author} {\bibfnamefont {I.~A.}\ \bibnamefont
  {Shovkovy}},\ }\bibfield  {title} {\bibinfo {title} {{Dimensional reduction
  and catalysis of dynamical symmetry breaking by a magnetic field}},\ }\href
  {https://doi.org/10.1016/0550-3213(96)00021-1} {\bibfield  {journal}
  {\bibinfo  {journal} {Nucl. Phys.}\ }\textbf {\bibinfo {volume} {B462}},\
  \bibinfo {pages} {249} (\bibinfo {year} {1996})},\ \Eprint
  {https://arxiv.org/abs/hep-ph/9509320} {arXiv:hep-ph/9509320 [hep-ph]}
  \BibitemShut {NoStop}%
%%CITATION = HEP-PH/9509320;%%
\bibitem [{\citenamefont {Miransky}\ and\ \citenamefont
  {Shovkovy}(2015)}]{Miransky:2015ava}%
  \BibitemOpen
  \bibfield  {author} {\bibinfo {author} {\bibfnamefont {V.~A.}\ \bibnamefont
  {Miransky}}\ and\ \bibinfo {author} {\bibfnamefont {I.~A.}\ \bibnamefont
  {Shovkovy}},\ }\bibfield  {title} {\bibinfo {title} {{Quantum field theory in
  a magnetic field: From quantum chromodynamics to graphene and Dirac
  semimetals}},\ }\href {https://doi.org/10.1016/j.physrep.2015.02.003}
  {\bibfield  {journal} {\bibinfo  {journal} {Phys. Rept.}\ }\textbf {\bibinfo
  {volume} {576}},\ \bibinfo {pages} {1} (\bibinfo {year} {2015})},\ \Eprint
  {https://arxiv.org/abs/1503.00732} {arXiv:1503.00732 [hep-ph]} \BibitemShut
  {NoStop}%
%%CITATION = ARXIV:1503.00732;%%
\bibitem [{\citenamefont {Bali}\ \emph
  {et~al.}(2012{\natexlab{a}})\citenamefont {Bali}, \citenamefont {Bruckmann},
  \citenamefont {Endrodi}, \citenamefont {Fodor}, \citenamefont {Katz},
  \citenamefont {Krieg}, \citenamefont {Schafer},\ and\ \citenamefont
  {Szabo}}]{Bali:2011qj}%
  \BibitemOpen
  \bibfield  {author} {\bibinfo {author} {\bibfnamefont {G.~S.}\ \bibnamefont
  {Bali}}, \bibinfo {author} {\bibfnamefont {F.}~\bibnamefont {Bruckmann}},
  \bibinfo {author} {\bibfnamefont {G.}~\bibnamefont {Endrodi}}, \bibinfo
  {author} {\bibfnamefont {Z.}~\bibnamefont {Fodor}}, \bibinfo {author}
  {\bibfnamefont {S.~D.}\ \bibnamefont {Katz}}, \bibinfo {author}
  {\bibfnamefont {S.}~\bibnamefont {Krieg}}, \bibinfo {author} {\bibfnamefont
  {A.}~\bibnamefont {Schafer}},\ and\ \bibinfo {author} {\bibfnamefont {K.~K.}\
  \bibnamefont {Szabo}},\ }\bibfield  {title} {\bibinfo {title} {{The QCD phase
  diagram for external magnetic fields}},\ }\href
  {https://doi.org/10.1007/JHEP02(2012)044} {\bibfield  {journal} {\bibinfo
  {journal} {JHEP}\ }\textbf {\bibinfo {volume} {02}},\ \bibinfo {pages}
  {044}},\ \Eprint {https://arxiv.org/abs/1111.4956} {arXiv:1111.4956
  [hep-lat]} \BibitemShut {NoStop}%
%%CITATION = ARXIV:1111.4956;%%
\bibitem [{\citenamefont {Bali}\ \emph
  {et~al.}(2012{\natexlab{b}})\citenamefont {Bali}, \citenamefont {Bruckmann},
  \citenamefont {Endrodi}, \citenamefont {Fodor}, \citenamefont {Katz},\ and\
  \citenamefont {Schafer}}]{Bali:2012zg}%
  \BibitemOpen
  \bibfield  {author} {\bibinfo {author} {\bibfnamefont {G.~S.}\ \bibnamefont
  {Bali}}, \bibinfo {author} {\bibfnamefont {F.}~\bibnamefont {Bruckmann}},
  \bibinfo {author} {\bibfnamefont {G.}~\bibnamefont {Endrodi}}, \bibinfo
  {author} {\bibfnamefont {Z.}~\bibnamefont {Fodor}}, \bibinfo {author}
  {\bibfnamefont {S.~D.}\ \bibnamefont {Katz}},\ and\ \bibinfo {author}
  {\bibfnamefont {A.}~\bibnamefont {Schafer}},\ }\bibfield  {title} {\bibinfo
  {title} {{QCD quark condensate in external magnetic fields}},\ }\href
  {https://doi.org/10.1103/PhysRevD.86.071502} {\bibfield  {journal} {\bibinfo
  {journal} {Phys. Rev.}\ }\textbf {\bibinfo {volume} {D86}},\ \bibinfo {pages}
  {071502} (\bibinfo {year} {2012}{\natexlab{b}})},\ \Eprint
  {https://arxiv.org/abs/1206.4205} {arXiv:1206.4205 [hep-lat]} \BibitemShut
  {NoStop}%
%%CITATION = ARXIV:1206.4205;%%
\bibitem [{\citenamefont {Preis}\ \emph {et~al.}(2011)\citenamefont {Preis},
  \citenamefont {Rebhan},\ and\ \citenamefont {Schmitt}}]{Preis:2010cq}%
  \BibitemOpen
  \bibfield  {author} {\bibinfo {author} {\bibfnamefont {F.}~\bibnamefont
  {Preis}}, \bibinfo {author} {\bibfnamefont {A.}~\bibnamefont {Rebhan}},\ and\
  \bibinfo {author} {\bibfnamefont {A.}~\bibnamefont {Schmitt}},\ }\bibfield
  {title} {\bibinfo {title} {{Inverse magnetic catalysis in dense holographic
  matter}},\ }\href {https://doi.org/10.1007/JHEP03(2011)033} {\bibfield
  {journal} {\bibinfo  {journal} {JHEP}\ }\textbf {\bibinfo {volume} {03}},\
  \bibinfo {pages} {033}},\ \Eprint {https://arxiv.org/abs/1012.4785}
  {arXiv:1012.4785 [hep-th]} \BibitemShut {NoStop}%
%%CITATION = ARXIV:1012.4785;%%
\bibitem [{\citenamefont {Fukushima}\ and\ \citenamefont
  {Warringa}(2008)}]{Fukushima:2007fc}%
  \BibitemOpen
  \bibfield  {author} {\bibinfo {author} {\bibfnamefont {K.}~\bibnamefont
  {Fukushima}}\ and\ \bibinfo {author} {\bibfnamefont {H.~J.}\ \bibnamefont
  {Warringa}},\ }\bibfield  {title} {\bibinfo {title} {{Color superconducting
  matter in a magnetic field}},\ }\href
  {https://doi.org/10.1103/PhysRevLett.100.032007} {\bibfield  {journal}
  {\bibinfo  {journal} {Phys. Rev. Lett.}\ }\textbf {\bibinfo {volume} {100}},\
  \bibinfo {pages} {032007} (\bibinfo {year} {2008})},\ \Eprint
  {https://arxiv.org/abs/0707.3785} {arXiv:0707.3785 [hep-ph]} \BibitemShut
  {NoStop}%
%%CITATION = ARXIV:0707.3785;%%
\bibitem [{\citenamefont {Ferrer}\ and\ \citenamefont {de~la
  Incera}(2007)}]{Ferrer:2007iw}%
  \BibitemOpen
  \bibfield  {author} {\bibinfo {author} {\bibfnamefont {E.~J.}\ \bibnamefont
  {Ferrer}}\ and\ \bibinfo {author} {\bibfnamefont {V.}~\bibnamefont {de~la
  Incera}},\ }\bibfield  {title} {\bibinfo {title} {{Magnetic Phases in
  Three-Flavor Color Superconductivity}},\ }\href
  {https://doi.org/10.1103/PhysRevD.76.045011} {\bibfield  {journal} {\bibinfo
  {journal} {Phys. Rev.}\ }\textbf {\bibinfo {volume} {D76}},\ \bibinfo {pages}
  {045011} (\bibinfo {year} {2007})},\ \Eprint
  {https://arxiv.org/abs/nucl-th/0703034} {arXiv:nucl-th/0703034 [nucl-th]}
  \BibitemShut {NoStop}%
%%CITATION = NUCL-TH/0703034;%%
\bibitem [{\citenamefont {Son}\ and\ \citenamefont
  {Stephanov}(2008)}]{Son:2007ny}%
  \BibitemOpen
  \bibfield  {author} {\bibinfo {author} {\bibfnamefont {D.~T.}\ \bibnamefont
  {Son}}\ and\ \bibinfo {author} {\bibfnamefont {M.~A.}\ \bibnamefont
  {Stephanov}},\ }\bibfield  {title} {\bibinfo {title} {{Axial anomaly and
  magnetism of nuclear and quark matter}},\ }\href
  {https://doi.org/10.1103/PhysRevD.77.014021} {\bibfield  {journal} {\bibinfo
  {journal} {Phys. Rev.}\ }\textbf {\bibinfo {volume} {D77}},\ \bibinfo {pages}
  {014021} (\bibinfo {year} {2008})},\ \Eprint
  {https://arxiv.org/abs/0710.1084} {arXiv:0710.1084 [hep-ph]} \BibitemShut
  {NoStop}%
%%CITATION = ARXIV:0710.1084;%%
\bibitem [{\citenamefont {Chernodub}(2011)}]{Chernodub:2011mc}%
  \BibitemOpen
  \bibfield  {author} {\bibinfo {author} {\bibfnamefont {M.~N.}\ \bibnamefont
  {Chernodub}},\ }\bibfield  {title} {\bibinfo {title} {{Spontaneous
  electromagnetic superconductivity of vacuum in strong magnetic field:
  evidence from the Nambu--Jona-Lasinio model}},\ }\href
  {https://doi.org/10.1103/PhysRevLett.106.142003} {\bibfield  {journal}
  {\bibinfo  {journal} {Phys. Rev. Lett.}\ }\textbf {\bibinfo {volume} {106}},\
  \bibinfo {pages} {142003} (\bibinfo {year} {2011})},\ \Eprint
  {https://arxiv.org/abs/1101.0117} {arXiv:1101.0117 [hep-ph]} \BibitemShut
  {NoStop}%
%%CITATION = ARXIV:1101.0117;%%
\bibitem [{\citenamefont {Hidaka}\ and\ \citenamefont
  {Yamamoto}(2013)}]{Hidaka:2012mz}%
  \BibitemOpen
  \bibfield  {author} {\bibinfo {author} {\bibfnamefont {Y.}~\bibnamefont
  {Hidaka}}\ and\ \bibinfo {author} {\bibfnamefont {A.}~\bibnamefont
  {Yamamoto}},\ }\bibfield  {title} {\bibinfo {title} {{Charged vector mesons
  in a strong magnetic field}},\ }\href
  {https://doi.org/10.1103/PhysRevD.87.094502} {\bibfield  {journal} {\bibinfo
  {journal} {Phys. Rev.}\ }\textbf {\bibinfo {volume} {D87}},\ \bibinfo {pages}
  {094502} (\bibinfo {year} {2013})},\ \Eprint
  {https://arxiv.org/abs/1209.0007} {arXiv:1209.0007 [hep-ph]} \BibitemShut
  {NoStop}%
%%CITATION = ARXIV:1209.0007;%%
\bibitem [{\citenamefont {Sinha}\ \emph {et~al.}(2013)\citenamefont {Sinha},
  \citenamefont {Huang},\ and\ \citenamefont {Sedrakian}}]{Sinha:2013dfa}%
  \BibitemOpen
  \bibfield  {author} {\bibinfo {author} {\bibfnamefont {M.}~\bibnamefont
  {Sinha}}, \bibinfo {author} {\bibfnamefont {X.-G.}\ \bibnamefont {Huang}},\
  and\ \bibinfo {author} {\bibfnamefont {A.}~\bibnamefont {Sedrakian}},\
  }\bibfield  {title} {\bibinfo {title} {{Strange quark matter in strong
  magnetic fields within a confining model}},\ }\href
  {https://doi.org/10.1103/PhysRevD.88.025008} {\bibfield  {journal} {\bibinfo
  {journal} {Phys. Rev.}\ }\textbf {\bibinfo {volume} {D88}},\ \bibinfo {pages}
  {025008} (\bibinfo {year} {2013})},\ \Eprint
  {https://arxiv.org/abs/1306.3300} {arXiv:1306.3300 [astro-ph.HE]}
  \BibitemShut {NoStop}%
%%CITATION = ARXIV:1306.3300;%%
\bibitem [{\citenamefont {Cao}\ and\ \citenamefont
  {Huang}(2016)}]{Cao:2015cka}%
  \BibitemOpen
  \bibfield  {author} {\bibinfo {author} {\bibfnamefont {G.}~\bibnamefont
  {Cao}}\ and\ \bibinfo {author} {\bibfnamefont {X.-G.}\ \bibnamefont
  {Huang}},\ }\bibfield  {title} {\bibinfo {title} {{Electromagnetic triangle
  anomaly and neutral pion condensation in QCD vacuum}},\ }\href
  {https://doi.org/10.1016/j.physletb.2016.03.066} {\bibfield  {journal}
  {\bibinfo  {journal} {Phys. Lett.}\ }\textbf {\bibinfo {volume} {B757}},\
  \bibinfo {pages} {1} (\bibinfo {year} {2016})},\ \Eprint
  {https://arxiv.org/abs/1509.06222} {arXiv:1509.06222 [hep-ph]} \BibitemShut
  {NoStop}%
%%CITATION = ARXIV:1509.06222;%%
\bibitem [{\citenamefont {Brauner}\ and\ \citenamefont
  {Yamamoto}(2017)}]{Brauner:2016pko}%
  \BibitemOpen
  \bibfield  {author} {\bibinfo {author} {\bibfnamefont {T.}~\bibnamefont
  {Brauner}}\ and\ \bibinfo {author} {\bibfnamefont {N.}~\bibnamefont
  {Yamamoto}},\ }\bibfield  {title} {\bibinfo {title} {{Chiral Soliton Lattice
  and Charged Pion Condensation in Strong Magnetic Fields}},\ }\href
  {https://doi.org/10.1007/JHEP04(2017)132} {\bibfield  {journal} {\bibinfo
  {journal} {JHEP}\ }\textbf {\bibinfo {volume} {04}},\ \bibinfo {pages}
  {132}},\ \Eprint {https://arxiv.org/abs/1609.05213} {arXiv:1609.05213
  [hep-ph]} \BibitemShut {NoStop}%
%%CITATION = ARXIV:1609.05213;%%
\bibitem [{\citenamefont {Hattori}\ \emph {et~al.}(2017)\citenamefont
  {Hattori}, \citenamefont {Itakura},\ and\ \citenamefont
  {Ozaki}}]{Hattori:2017qio}%
  \BibitemOpen
  \bibfield  {author} {\bibinfo {author} {\bibfnamefont {K.}~\bibnamefont
  {Hattori}}, \bibinfo {author} {\bibfnamefont {K.}~\bibnamefont {Itakura}},\
  and\ \bibinfo {author} {\bibfnamefont {S.}~\bibnamefont {Ozaki}},\ }\bibfield
   {title} {\bibinfo {title} {{Anatomy of the magnetic catalysis by
  renormalization-group method}},\ }\href
  {https://doi.org/10.1016/j.physletb.2017.11.004} {\bibfield  {journal}
  {\bibinfo  {journal} {Phys. Lett.}\ }\textbf {\bibinfo {volume} {B775}},\
  \bibinfo {pages} {283} (\bibinfo {year} {2017})},\ \Eprint
  {https://arxiv.org/abs/1706.04913} {arXiv:1706.04913 [hep-ph]} \BibitemShut
  {NoStop}%
%%CITATION = ARXIV:1706.04913;%%
\bibitem [{\citenamefont {Ozaki}\ \emph {et~al.}(2016)\citenamefont {Ozaki},
  \citenamefont {Itakura},\ and\ \citenamefont {Kuramoto}}]{Ozaki:2015sya}%
  \BibitemOpen
  \bibfield  {author} {\bibinfo {author} {\bibfnamefont {S.}~\bibnamefont
  {Ozaki}}, \bibinfo {author} {\bibfnamefont {K.}~\bibnamefont {Itakura}},\
  and\ \bibinfo {author} {\bibfnamefont {Y.}~\bibnamefont {Kuramoto}},\
  }\bibfield  {title} {\bibinfo {title} {{Magnetically induced QCD Kondo
  effect}},\ }\href {https://doi.org/10.1103/PhysRevD.94.074013} {\bibfield
  {journal} {\bibinfo  {journal} {Phys. Rev.}\ }\textbf {\bibinfo {volume}
  {D94}},\ \bibinfo {pages} {074013} (\bibinfo {year} {2016})},\ \Eprint
  {https://arxiv.org/abs/1509.06966} {arXiv:1509.06966 [hep-ph]} \BibitemShut
  {NoStop}%
%%CITATION = ARXIV:1509.06966;%%
\bibitem [{\citenamefont {Chen}\ \emph {et~al.}(2017)\citenamefont {Chen},
  \citenamefont {Fukushima}, \citenamefont {Huang},\ and\ \citenamefont
  {Mameda}}]{Chen:2017xrj}%
  \BibitemOpen
  \bibfield  {author} {\bibinfo {author} {\bibfnamefont {H.-L.}\ \bibnamefont
  {Chen}}, \bibinfo {author} {\bibfnamefont {K.}~\bibnamefont {Fukushima}},
  \bibinfo {author} {\bibfnamefont {X.-G.}\ \bibnamefont {Huang}},\ and\
  \bibinfo {author} {\bibfnamefont {K.}~\bibnamefont {Mameda}},\ }\bibfield
  {title} {\bibinfo {title} {{Surface Magnetic Catalysis}},\ }\href
  {https://doi.org/10.1103/PhysRevD.96.054032} {\bibfield  {journal} {\bibinfo
  {journal} {Phys. Rev.}\ }\textbf {\bibinfo {volume} {D96}},\ \bibinfo {pages}
  {054032} (\bibinfo {year} {2017})},\ \Eprint
  {https://arxiv.org/abs/1707.09130} {arXiv:1707.09130 [hep-ph]} \BibitemShut
  {NoStop}%
%%CITATION = ARXIV:1707.09130;%%
\bibitem [{\citenamefont {Deng}\ and\ \citenamefont
  {Huang}(2016)}]{Deng:2016gyh}%
  \BibitemOpen
  \bibfield  {author} {\bibinfo {author} {\bibfnamefont {W.-T.}\ \bibnamefont
  {Deng}}\ and\ \bibinfo {author} {\bibfnamefont {X.-G.}\ \bibnamefont
  {Huang}},\ }\bibfield  {title} {\bibinfo {title} {{Vorticity in Heavy-Ion
  Collisions}},\ }\href {https://doi.org/10.1103/PhysRevC.93.064907} {\bibfield
   {journal} {\bibinfo  {journal} {Phys. Rev.}\ }\textbf {\bibinfo {volume}
  {C93}},\ \bibinfo {pages} {064907} (\bibinfo {year} {2016})},\ \Eprint
  {https://arxiv.org/abs/1603.06117} {arXiv:1603.06117 [nucl-th]} \BibitemShut
  {NoStop}%
%%CITATION = ARXIV:1603.06117;%%
\bibitem [{\citenamefont {Jiang}\ \emph {et~al.}(2016)\citenamefont {Jiang},
  \citenamefont {Lin},\ and\ \citenamefont {Liao}}]{Jiang:2016woz}%
  \BibitemOpen
  \bibfield  {author} {\bibinfo {author} {\bibfnamefont {Y.}~\bibnamefont
  {Jiang}}, \bibinfo {author} {\bibfnamefont {Z.-W.}\ \bibnamefont {Lin}},\
  and\ \bibinfo {author} {\bibfnamefont {J.}~\bibnamefont {Liao}},\ }\bibfield
  {title} {\bibinfo {title} {{Rotating quark-gluon plasma in relativistic heavy
  ion collisions}},\ }\href {https://doi.org/10.1103/PhysRevC.94.044910,
  10.1103/PhysRevC.95.049904} {\bibfield  {journal} {\bibinfo  {journal} {Phys.
  Rev.}\ }\textbf {\bibinfo {volume} {C94}},\ \bibinfo {pages} {044910}
  (\bibinfo {year} {2016})},\ \bibinfo {note} {[Erratum: Phys.
  Rev.C95,no.4,049904(2017)]},\ \Eprint {https://arxiv.org/abs/1602.06580}
  {arXiv:1602.06580 [hep-ph]} \BibitemShut {NoStop}%
%%CITATION = ARXIV:1602.06580;%%
\bibitem [{\citenamefont {Wei}\ \emph {et~al.}(2019)\citenamefont {Wei},
  \citenamefont {Deng},\ and\ \citenamefont {Huang}}]{Wei:2018zfb}%
  \BibitemOpen
  \bibfield  {author} {\bibinfo {author} {\bibfnamefont {D.-X.}\ \bibnamefont
  {Wei}}, \bibinfo {author} {\bibfnamefont {W.-T.}\ \bibnamefont {Deng}},\ and\
  \bibinfo {author} {\bibfnamefont {X.-G.}\ \bibnamefont {Huang}},\ }\bibfield
  {title} {\bibinfo {title} {{Thermal vorticity and spin polarization in
  heavy-ion collisions}},\ }\href {https://doi.org/10.1103/PhysRevC.99.014905}
  {\bibfield  {journal} {\bibinfo  {journal} {Phys. Rev.}\ }\textbf {\bibinfo
  {volume} {C99}},\ \bibinfo {pages} {014905} (\bibinfo {year} {2019})},\
  \Eprint {https://arxiv.org/abs/1810.00151} {arXiv:1810.00151 [nucl-th]}
  \BibitemShut {NoStop}%
%%CITATION = ARXIV:1810.00151;%%
\bibitem [{\citenamefont {Liang}\ and\ \citenamefont
  {Wang}(2005)}]{Liang:2004ph}%
  \BibitemOpen
  \bibfield  {author} {\bibinfo {author} {\bibfnamefont {Z.-T.}\ \bibnamefont
  {Liang}}\ and\ \bibinfo {author} {\bibfnamefont {X.-N.}\ \bibnamefont
  {Wang}},\ }\bibfield  {title} {\bibinfo {title} {{Globally polarized
  quark-gluon plasma in non-central A+A collisions}},\ }\href
  {https://doi.org/10.1103/PhysRevLett.94.102301,
  10.1103/PhysRevLett.96.039901} {\bibfield  {journal} {\bibinfo  {journal}
  {Phys. Rev. Lett.}\ }\textbf {\bibinfo {volume} {94}},\ \bibinfo {pages}
  {102301} (\bibinfo {year} {2005})},\ \bibinfo {note} {[Erratum: Phys. Rev.
  Lett.96,039901(2006)]},\ \Eprint {https://arxiv.org/abs/nucl-th/0410079}
  {arXiv:nucl-th/0410079 [nucl-th]} \BibitemShut {NoStop}%
%%CITATION = NUCL-TH/0410079;%%
\bibitem [{\citenamefont {Adamczyk}\ \emph {et~al.}(2017)\citenamefont
  {Adamczyk} \emph {et~al.}}]{STAR:2017ckg}%
  \BibitemOpen
  \bibfield  {author} {\bibinfo {author} {\bibfnamefont {L.}~\bibnamefont
  {Adamczyk}} \emph {et~al.} (\bibinfo {collaboration} {STAR}),\ }\bibfield
  {title} {\bibinfo {title} {{Global $\Lambda$ hyperon polarization in nuclear
  collisions: evidence for the most vortical fluid}},\ }\href
  {https://doi.org/10.1038/nature23004} {\bibfield  {journal} {\bibinfo
  {journal} {Nature}\ }\textbf {\bibinfo {volume} {548}},\ \bibinfo {pages}
  {62} (\bibinfo {year} {2017})},\ \Eprint {https://arxiv.org/abs/1701.06657}
  {arXiv:1701.06657 [nucl-ex]} \BibitemShut {NoStop}%
%%CITATION = ARXIV:1701.06657;%%
\bibitem [{\citenamefont {Vilenkin}(1979)}]{Vilenkin:1979ui}%
  \BibitemOpen
  \bibfield  {author} {\bibinfo {author} {\bibfnamefont {A.}~\bibnamefont
  {Vilenkin}},\ }\bibfield  {title} {\bibinfo {title} {Macroscopic parity
  violating effects: Neutrino fluxes from rotating balck holes and in rotating
  thermal radiation},\ }\href {https://doi.org/10.1103/PhysRevD.20.1807}
  {\bibfield  {journal} {\bibinfo  {journal} {Phys. Rev.}\ }\textbf {\bibinfo
  {volume} {D20}},\ \bibinfo {pages} {1807} (\bibinfo {year}
  {1979})}\BibitemShut {NoStop}%
%%CITATION = PHRVA,D20,1807;%%
\bibitem [{\citenamefont {Vilenkin}(1980)}]{Vilenkin:1980zv}%
  \BibitemOpen
  \bibfield  {author} {\bibinfo {author} {\bibfnamefont {A.}~\bibnamefont
  {Vilenkin}},\ }\bibfield  {title} {\bibinfo {title} {{Quantum field theory at
  finite temperature in a rotating system}},\ }\href
  {https://doi.org/10.1103/PhysRevD.21.2260} {\bibfield  {journal} {\bibinfo
  {journal} {Phys. Rev.}\ }\textbf {\bibinfo {volume} {D21}},\ \bibinfo {pages}
  {2260} (\bibinfo {year} {1980})}\BibitemShut {NoStop}%
%%CITATION = PHRVA,D21,2260;%%
\bibitem [{\citenamefont {Son}\ and\ \citenamefont
  {Surowka}(2009)}]{Son:2009tf}%
  \BibitemOpen
  \bibfield  {author} {\bibinfo {author} {\bibfnamefont {D.~T.}\ \bibnamefont
  {Son}}\ and\ \bibinfo {author} {\bibfnamefont {P.}~\bibnamefont {Surowka}},\
  }\bibfield  {title} {\bibinfo {title} {{Hydrodynamics with Triangle
  Anomalies}},\ }\href {https://doi.org/10.1103/PhysRevLett.103.191601}
  {\bibfield  {journal} {\bibinfo  {journal} {Phys. Rev. Lett.}\ }\textbf
  {\bibinfo {volume} {103}},\ \bibinfo {pages} {191601} (\bibinfo {year}
  {2009})},\ \Eprint {https://arxiv.org/abs/0906.5044} {arXiv:0906.5044
  [hep-th]} \BibitemShut {NoStop}%
%%CITATION = ARXIV:0906.5044;%%
\bibitem [{\citenamefont {Liu}\ \emph {et~al.}(2019)\citenamefont {Liu},
  \citenamefont {Gao}, \citenamefont {Mameda},\ and\ \citenamefont
  {Huang}}]{Liu:2018xip}%
  \BibitemOpen
  \bibfield  {author} {\bibinfo {author} {\bibfnamefont {Y.-C.}\ \bibnamefont
  {Liu}}, \bibinfo {author} {\bibfnamefont {L.-L.}\ \bibnamefont {Gao}},
  \bibinfo {author} {\bibfnamefont {K.}~\bibnamefont {Mameda}},\ and\ \bibinfo
  {author} {\bibfnamefont {X.-G.}\ \bibnamefont {Huang}},\ }\bibfield  {title}
  {\bibinfo {title} {{Chiral kinetic theory in curved spacetime}},\ }\href
  {https://doi.org/10.1103/PhysRevD.99.085014} {\bibfield  {journal} {\bibinfo
  {journal} {Phys. Rev.}\ }\textbf {\bibinfo {volume} {D99}},\ \bibinfo {pages}
  {085014} (\bibinfo {year} {2019})},\ \Eprint
  {https://arxiv.org/abs/1812.10127} {arXiv:1812.10127 [hep-th]} \BibitemShut
  {NoStop}%
%%CITATION = ARXIV:1812.10127;%%
\bibitem [{\citenamefont {Chen}\ \emph {et~al.}(2016)\citenamefont {Chen},
  \citenamefont {Fukushima}, \citenamefont {Huang},\ and\ \citenamefont
  {Mameda}}]{Chen:2015hfc}%
  \BibitemOpen
  \bibfield  {author} {\bibinfo {author} {\bibfnamefont {H.-L.}\ \bibnamefont
  {Chen}}, \bibinfo {author} {\bibfnamefont {K.}~\bibnamefont {Fukushima}},
  \bibinfo {author} {\bibfnamefont {X.-G.}\ \bibnamefont {Huang}},\ and\
  \bibinfo {author} {\bibfnamefont {K.}~\bibnamefont {Mameda}},\ }\bibfield
  {title} {\bibinfo {title} {{Analogy between rotation and density for Dirac
  fermions in a magnetic field}},\ }\href
  {https://doi.org/10.1103/PhysRevD.93.104052} {\bibfield  {journal} {\bibinfo
  {journal} {Phys. Rev.}\ }\textbf {\bibinfo {volume} {D93}},\ \bibinfo {pages}
  {104052} (\bibinfo {year} {2016})},\ \Eprint
  {https://arxiv.org/abs/1512.08974} {arXiv:1512.08974 [hep-ph]} \BibitemShut
  {NoStop}%
%%CITATION = ARXIV:1512.08974;%%
\bibitem [{\citenamefont {Jiang}\ and\ \citenamefont
  {Liao}(2016)}]{Jiang:2016wvv}%
  \BibitemOpen
  \bibfield  {author} {\bibinfo {author} {\bibfnamefont {Y.}~\bibnamefont
  {Jiang}}\ and\ \bibinfo {author} {\bibfnamefont {J.}~\bibnamefont {Liao}},\
  }\bibfield  {title} {\bibinfo {title} {{Pairing Phase Transitions of Matter
  under Rotation}},\ }\href {https://doi.org/10.1103/PhysRevLett.117.192302}
  {\bibfield  {journal} {\bibinfo  {journal} {Phys. Rev. Lett.}\ }\textbf
  {\bibinfo {volume} {117}},\ \bibinfo {pages} {192302} (\bibinfo {year}
  {2016})},\ \Eprint {https://arxiv.org/abs/1606.03808} {arXiv:1606.03808
  [hep-ph]} \BibitemShut {NoStop}%
%%CITATION = ARXIV:1606.03808;%%
\bibitem [{\citenamefont {Ebihara}\ \emph {et~al.}(2017)\citenamefont
  {Ebihara}, \citenamefont {Fukushima},\ and\ \citenamefont
  {Mameda}}]{Ebihara:2016fwa}%
  \BibitemOpen
  \bibfield  {author} {\bibinfo {author} {\bibfnamefont {S.}~\bibnamefont
  {Ebihara}}, \bibinfo {author} {\bibfnamefont {K.}~\bibnamefont {Fukushima}},\
  and\ \bibinfo {author} {\bibfnamefont {K.}~\bibnamefont {Mameda}},\
  }\bibfield  {title} {\bibinfo {title} {{Boundary effects and gapped
  dispersion in rotating fermionic matter}},\ }\href
  {https://doi.org/10.1016/j.physletb.2016.11.010} {\bibfield  {journal}
  {\bibinfo  {journal} {Phys. Lett.}\ }\textbf {\bibinfo {volume} {B764}},\
  \bibinfo {pages} {94} (\bibinfo {year} {2017})},\ \Eprint
  {https://arxiv.org/abs/1608.00336} {arXiv:1608.00336 [hep-ph]} \BibitemShut
  {NoStop}%
%%CITATION = ARXIV:1608.00336;%%
\bibitem [{\citenamefont {Chernodub}\ and\ \citenamefont
  {Gongyo}(2017{\natexlab{a}})}]{Chernodub:2016kxh}%
  \BibitemOpen
  \bibfield  {author} {\bibinfo {author} {\bibfnamefont {M.~N.}\ \bibnamefont
  {Chernodub}}\ and\ \bibinfo {author} {\bibfnamefont {S.}~\bibnamefont
  {Gongyo}},\ }\bibfield  {title} {\bibinfo {title} {{Interacting fermions in
  rotation: chiral symmetry restoration, moment of inertia and
  thermodynamics}},\ }\href {https://doi.org/10.1007/JHEP01(2017)136}
  {\bibfield  {journal} {\bibinfo  {journal} {JHEP}\ }\textbf {\bibinfo
  {volume} {01}},\ \bibinfo {pages} {136}},\ \Eprint
  {https://arxiv.org/abs/1611.02598} {arXiv:1611.02598 [hep-th]} \BibitemShut
  {NoStop}%
%%CITATION = ARXIV:1611.02598;%%
\bibitem [{\citenamefont {Huang}\ \emph {et~al.}(2018)\citenamefont {Huang},
  \citenamefont {Nishimura},\ and\ \citenamefont {Yamamoto}}]{Huang:2017pqe}%
  \BibitemOpen
  \bibfield  {author} {\bibinfo {author} {\bibfnamefont {X.-G.}\ \bibnamefont
  {Huang}}, \bibinfo {author} {\bibfnamefont {K.}~\bibnamefont {Nishimura}},\
  and\ \bibinfo {author} {\bibfnamefont {N.}~\bibnamefont {Yamamoto}},\
  }\bibfield  {title} {\bibinfo {title} {{Anomalous effects of dense matter
  under rotation}},\ }\href {https://doi.org/10.1007/JHEP02(2018)069}
  {\bibfield  {journal} {\bibinfo  {journal} {JHEP}\ }\textbf {\bibinfo
  {volume} {02}},\ \bibinfo {pages} {069}},\ \Eprint
  {https://arxiv.org/abs/1711.02190} {arXiv:1711.02190 [hep-ph]} \BibitemShut
  {NoStop}%
%%CITATION = ARXIV:1711.02190;%%
\bibitem [{\citenamefont {Chernodub}\ and\ \citenamefont
  {Gongyo}(2017{\natexlab{b}})}]{Chernodub:2017ref}%
  \BibitemOpen
  \bibfield  {author} {\bibinfo {author} {\bibfnamefont {M.~N.}\ \bibnamefont
  {Chernodub}}\ and\ \bibinfo {author} {\bibfnamefont {S.}~\bibnamefont
  {Gongyo}},\ }\bibfield  {title} {\bibinfo {title} {{Effects of rotation and
  boundaries on chiral symmetry breaking of relativistic fermions}},\ }\href
  {https://doi.org/10.1103/PhysRevD.95.096006} {\bibfield  {journal} {\bibinfo
  {journal} {Phys. Rev.}\ }\textbf {\bibinfo {volume} {D95}},\ \bibinfo {pages}
  {096006} (\bibinfo {year} {2017}{\natexlab{b}})},\ \Eprint
  {https://arxiv.org/abs/1702.08266} {arXiv:1702.08266 [hep-th]} \BibitemShut
  {NoStop}%
%%CITATION = ARXIV:1702.08266;%%
\bibitem [{\citenamefont {Chernodub}\ and\ \citenamefont
  {Gongyo}(2017{\natexlab{c}})}]{Chernodub:2017mvp}%
  \BibitemOpen
  \bibfield  {author} {\bibinfo {author} {\bibfnamefont {M.~N.}\ \bibnamefont
  {Chernodub}}\ and\ \bibinfo {author} {\bibfnamefont {S.}~\bibnamefont
  {Gongyo}},\ }\bibfield  {title} {\bibinfo {title} {{Edge states and
  thermodynamics of rotating relativistic fermions under magnetic field}},\
  }\href {https://doi.org/10.1103/PhysRevD.96.096014} {\bibfield  {journal}
  {\bibinfo  {journal} {Phys. Rev.}\ }\textbf {\bibinfo {volume} {D96}},\
  \bibinfo {pages} {096014} (\bibinfo {year} {2017}{\natexlab{c}})},\ \Eprint
  {https://arxiv.org/abs/1706.08448} {arXiv:1706.08448 [hep-th]} \BibitemShut
  {NoStop}%
%%CITATION = ARXIV:1706.08448;%%
\bibitem [{\citenamefont {Liu}\ and\ \citenamefont
  {Zahed}(2018{\natexlab{a}})}]{Liu:2017spl}%
  \BibitemOpen
  \bibfield  {author} {\bibinfo {author} {\bibfnamefont {Y.}~\bibnamefont
  {Liu}}\ and\ \bibinfo {author} {\bibfnamefont {I.}~\bibnamefont {Zahed}},\
  }\bibfield  {title} {\bibinfo {title} {{Pion Condensation by Rotation in a
  Magnetic field}},\ }\href {https://doi.org/10.1103/PhysRevLett.120.032001}
  {\bibfield  {journal} {\bibinfo  {journal} {Phys. Rev. Lett.}\ }\textbf
  {\bibinfo {volume} {120}},\ \bibinfo {pages} {032001} (\bibinfo {year}
  {2018}{\natexlab{a}})},\ \Eprint {https://arxiv.org/abs/1711.08354}
  {arXiv:1711.08354 [hep-ph]} \BibitemShut {NoStop}%
%%CITATION = ARXIV:1711.08354;%%
\bibitem [{\citenamefont {Wang}\ \emph
  {et~al.}(2019{\natexlab{a}})\citenamefont {Wang}, \citenamefont {Jiang},
  \citenamefont {He},\ and\ \citenamefont {Zhuang}}]{Wang:2018zrn}%
  \BibitemOpen
  \bibfield  {author} {\bibinfo {author} {\bibfnamefont {L.}~\bibnamefont
  {Wang}}, \bibinfo {author} {\bibfnamefont {Y.}~\bibnamefont {Jiang}},
  \bibinfo {author} {\bibfnamefont {L.}~\bibnamefont {He}},\ and\ \bibinfo
  {author} {\bibfnamefont {P.}~\bibnamefont {Zhuang}},\ }\bibfield  {title}
  {\bibinfo {title} {{Local suppression and enhancement of the pairing
  condensate under rotation}},\ }\href
  {https://doi.org/10.1103/PhysRevC.100.034902} {\bibfield  {journal} {\bibinfo
   {journal} {Phys. Rev.}\ }\textbf {\bibinfo {volume} {C100}},\ \bibinfo
  {pages} {034902} (\bibinfo {year} {2019}{\natexlab{a}})},\ \Eprint
  {https://arxiv.org/abs/1901.00804} {arXiv:1901.00804 [nucl-th]} \BibitemShut
  {NoStop}%
%%CITATION = ARXIV:1901.00804;%%
\bibitem [{\citenamefont {Zhang}\ \emph {et~al.}(2020)\citenamefont {Zhang},
  \citenamefont {Hou},\ and\ \citenamefont {Liao}}]{Zhang:2018ome}%
  \BibitemOpen
  \bibfield  {author} {\bibinfo {author} {\bibfnamefont {H.}~\bibnamefont
  {Zhang}}, \bibinfo {author} {\bibfnamefont {D.}~\bibnamefont {Hou}},\ and\
  \bibinfo {author} {\bibfnamefont {J.}~\bibnamefont {Liao}},\ }\bibfield
  {title} {\bibinfo {title} {{Mesonic Condensation in Isospin Matter under
  Rotation}},\ }\href {https://doi.org/10.1088/1674-1137/abae4d} {\bibfield
  {journal} {\bibinfo  {journal} {Chin. Phys. C}\ }\textbf {\bibinfo {volume}
  {44}},\ \bibinfo {pages} {111001} (\bibinfo {year} {2020})},\ \Eprint
  {https://arxiv.org/abs/1812.11787} {arXiv:1812.11787 [hep-ph]} \BibitemShut
  {NoStop}%
\bibitem [{\citenamefont {Wang}\ \emph
  {et~al.}(2019{\natexlab{b}})\citenamefont {Wang}, \citenamefont {Wei},
  \citenamefont {Li},\ and\ \citenamefont {Huang}}]{Wang:2018sur}%
  \BibitemOpen
  \bibfield  {author} {\bibinfo {author} {\bibfnamefont {X.}~\bibnamefont
  {Wang}}, \bibinfo {author} {\bibfnamefont {M.}~\bibnamefont {Wei}}, \bibinfo
  {author} {\bibfnamefont {Z.}~\bibnamefont {Li}},\ and\ \bibinfo {author}
  {\bibfnamefont {M.}~\bibnamefont {Huang}},\ }\bibfield  {title} {\bibinfo
  {title} {{Quark matter under rotation in the NJL model with vector
  interaction}},\ }\href {https://doi.org/10.1103/PhysRevD.99.016018}
  {\bibfield  {journal} {\bibinfo  {journal} {Phys. Rev.}\ }\textbf {\bibinfo
  {volume} {D99}},\ \bibinfo {pages} {016018} (\bibinfo {year}
  {2019}{\natexlab{b}})},\ \Eprint {https://arxiv.org/abs/1808.01931}
  {arXiv:1808.01931 [hep-ph]} \BibitemShut {NoStop}%
%%CITATION = ARXIV:1808.01931;%%
\bibitem [{\citenamefont {Wang}\ \emph
  {et~al.}(2019{\natexlab{c}})\citenamefont {Wang}, \citenamefont {Jiang},
  \citenamefont {He},\ and\ \citenamefont {Zhuang}}]{Wang:2019nhd}%
  \BibitemOpen
  \bibfield  {author} {\bibinfo {author} {\bibfnamefont {L.}~\bibnamefont
  {Wang}}, \bibinfo {author} {\bibfnamefont {Y.}~\bibnamefont {Jiang}},
  \bibinfo {author} {\bibfnamefont {L.}~\bibnamefont {He}},\ and\ \bibinfo
  {author} {\bibfnamefont {P.}~\bibnamefont {Zhuang}},\ }\bibfield  {title}
  {\bibinfo {title} {{Chiral vortices and pseudoscalar condensation due to
  rotation}},\ }\href {https://doi.org/10.1103/PhysRevD.100.114009} {\bibfield
  {journal} {\bibinfo  {journal} {Phys. Rev. D}\ }\textbf {\bibinfo {volume}
  {100}},\ \bibinfo {pages} {114009} (\bibinfo {year} {2019}{\natexlab{c}})},\
  \Eprint {https://arxiv.org/abs/1901.04697} {arXiv:1901.04697 [nucl-th]}
  \BibitemShut {NoStop}%
\bibitem [{\citenamefont {Liu}\ and\ \citenamefont
  {Zahed}(2018{\natexlab{b}})}]{Liu:2017zhl}%
  \BibitemOpen
  \bibfield  {author} {\bibinfo {author} {\bibfnamefont {Y.}~\bibnamefont
  {Liu}}\ and\ \bibinfo {author} {\bibfnamefont {I.}~\bibnamefont {Zahed}},\
  }\bibfield  {title} {\bibinfo {title} {{Rotating Dirac fermions in a magnetic
  field in 1+2 and 1+3 dimensions}},\ }\href
  {https://doi.org/10.1103/PhysRevD.98.014017} {\bibfield  {journal} {\bibinfo
  {journal} {Phys. Rev.}\ }\textbf {\bibinfo {volume} {D98}},\ \bibinfo {pages}
  {014017} (\bibinfo {year} {2018}{\natexlab{b}})},\ \Eprint
  {https://arxiv.org/abs/1710.02895} {arXiv:1710.02895 [hep-ph]} \BibitemShut
  {NoStop}%
%%CITATION = ARXIV:1710.02895;%%
\bibitem [{\citenamefont {Son}\ and\ \citenamefont
  {Stephanov}(2001)}]{Son:2000xc}%
  \BibitemOpen
  \bibfield  {author} {\bibinfo {author} {\bibfnamefont {D.~T.}\ \bibnamefont
  {Son}}\ and\ \bibinfo {author} {\bibfnamefont {M.~A.}\ \bibnamefont
  {Stephanov}},\ }\bibfield  {title} {\bibinfo {title} {{QCD at finite isospin
  density}},\ }\href {https://doi.org/10.1103/PhysRevLett.86.592} {\bibfield
  {journal} {\bibinfo  {journal} {Phys. Rev. Lett.}\ }\textbf {\bibinfo
  {volume} {86}},\ \bibinfo {pages} {592} (\bibinfo {year} {2001})},\ \Eprint
  {https://arxiv.org/abs/hep-ph/0005225} {arXiv:hep-ph/0005225 [hep-ph]}
  \BibitemShut {NoStop}%
%%CITATION = HEP-PH/0005225;%%
\bibitem [{\citenamefont {He}\ \emph {et~al.}(2005)\citenamefont {He},
  \citenamefont {Jin},\ and\ \citenamefont {Zhuang}}]{He:2005nk}%
  \BibitemOpen
  \bibfield  {author} {\bibinfo {author} {\bibfnamefont {L.-Y.}\ \bibnamefont
  {He}}, \bibinfo {author} {\bibfnamefont {M.}~\bibnamefont {Jin}},\ and\
  \bibinfo {author} {\bibfnamefont {P.-F.}\ \bibnamefont {Zhuang}},\ }\bibfield
   {title} {\bibinfo {title} {{Pion superfluidity and meson properties at
  finite isospin density}},\ }\href
  {https://doi.org/10.1103/PhysRevD.71.116001} {\bibfield  {journal} {\bibinfo
  {journal} {Phys. Rev.}\ }\textbf {\bibinfo {volume} {D71}},\ \bibinfo {pages}
  {116001} (\bibinfo {year} {2005})},\ \Eprint
  {https://arxiv.org/abs/hep-ph/0503272} {arXiv:hep-ph/0503272 [hep-ph]}
  \BibitemShut {NoStop}%
%%CITATION = HEP-PH/0503272;%%
\bibitem [{\citenamefont {Sun}\ \emph {et~al.}(2007)\citenamefont {Sun},
  \citenamefont {He},\ and\ \citenamefont {Zhuang}}]{Sun:2007fc}%
  \BibitemOpen
  \bibfield  {author} {\bibinfo {author} {\bibfnamefont {G.-F.}\ \bibnamefont
  {Sun}}, \bibinfo {author} {\bibfnamefont {L.}~\bibnamefont {He}},\ and\
  \bibinfo {author} {\bibfnamefont {P.}~\bibnamefont {Zhuang}},\ }\bibfield
  {title} {\bibinfo {title} {{BEC-BCS crossover in the Nambu-Jona-Lasinio model
  of QCD}},\ }\href {https://doi.org/10.1103/PhysRevD.75.096004} {\bibfield
  {journal} {\bibinfo  {journal} {Phys. Rev.}\ }\textbf {\bibinfo {volume}
  {D75}},\ \bibinfo {pages} {096004} (\bibinfo {year} {2007})},\ \Eprint
  {https://arxiv.org/abs/hep-ph/0703159} {arXiv:hep-ph/0703159 [hep-ph]}
  \BibitemShut {NoStop}%
%%CITATION = HEP-PH/0703159;%%
\bibitem [{\citenamefont {He}\ \emph {et~al.}(2006)\citenamefont {He},
  \citenamefont {Jin},\ and\ \citenamefont {Zhuang}}]{He:2006tn}%
  \BibitemOpen
  \bibfield  {author} {\bibinfo {author} {\bibfnamefont {L.}~\bibnamefont
  {He}}, \bibinfo {author} {\bibfnamefont {M.}~\bibnamefont {Jin}},\ and\
  \bibinfo {author} {\bibfnamefont {P.}~\bibnamefont {Zhuang}},\ }\bibfield
  {title} {\bibinfo {title} {{Pion Condensation in Baryonic Matter: from Sarma
  Phase to Larkin-Ovchinnikov-Fudde-Ferrell Phase}},\ }\href
  {https://doi.org/10.1103/PhysRevD.74.036005} {\bibfield  {journal} {\bibinfo
  {journal} {Phys. Rev.}\ }\textbf {\bibinfo {volume} {D74}},\ \bibinfo {pages}
  {036005} (\bibinfo {year} {2006})},\ \Eprint
  {https://arxiv.org/abs/hep-ph/0604224} {arXiv:hep-ph/0604224 [hep-ph]}
  \BibitemShut {NoStop}%
%%CITATION = HEP-PH/0604224;%%
\bibitem [{\citenamefont {Guo}\ \emph {et~al.}(2022)\citenamefont {Guo},
  \citenamefont {Li}, \citenamefont {Mu},\ and\ \citenamefont
  {He}}]{Guo:2021gbz}%
  \BibitemOpen
  \bibfield  {author} {\bibinfo {author} {\bibfnamefont {T.}~\bibnamefont
  {Guo}}, \bibinfo {author} {\bibfnamefont {J.}~\bibnamefont {Li}}, \bibinfo
  {author} {\bibfnamefont {C.}~\bibnamefont {Mu}},\ and\ \bibinfo {author}
  {\bibfnamefont {L.}~\bibnamefont {He}},\ }\bibfield  {title} {\bibinfo
  {title} {{Formation of a supergiant quantum vortex in a relativistic
  Bose-Einstein condensate driven by rotation and a parallel magnetic field}},\
  }\href {https://doi.org/10.1103/PhysRevD.106.094010} {\bibfield  {journal}
  {\bibinfo  {journal} {Phys. Rev. D}\ }\textbf {\bibinfo {volume} {106}},\
  \bibinfo {pages} {094010} (\bibinfo {year} {2022})},\ \Eprint
  {https://arxiv.org/abs/2111.13159} {arXiv:2111.13159 [nucl-th]} \BibitemShut
  {NoStop}%
\bibitem [{\citenamefont {Hattori}\ and\ \citenamefont
  {Yin}(2016)}]{Hattori:2016njk}%
  \BibitemOpen
  \bibfield  {author} {\bibinfo {author} {\bibfnamefont {K.}~\bibnamefont
  {Hattori}}\ and\ \bibinfo {author} {\bibfnamefont {Y.}~\bibnamefont {Yin}},\
  }\bibfield  {title} {\bibinfo {title} {{Charge redistribution from anomalous
  magnetovorticity coupling}},\ }\href
  {https://doi.org/10.1103/PhysRevLett.117.152002} {\bibfield  {journal}
  {\bibinfo  {journal} {Phys. Rev. Lett.}\ }\textbf {\bibinfo {volume} {117}},\
  \bibinfo {pages} {152002} (\bibinfo {year} {2016})},\ \Eprint
  {https://arxiv.org/abs/1607.01513} {arXiv:1607.01513 [hep-th]} \BibitemShut
  {NoStop}%
%%CITATION = ARXIV:1607.01513;%%
\bibitem [{\citenamefont {Poisson}\ \emph {et~al.}(2011)\citenamefont
  {Poisson}, \citenamefont {Pound},\ and\ \citenamefont
  {Vega}}]{Poisson:2011nh}%
  \BibitemOpen
  \bibfield  {author} {\bibinfo {author} {\bibfnamefont {E.}~\bibnamefont
  {Poisson}}, \bibinfo {author} {\bibfnamefont {A.}~\bibnamefont {Pound}},\
  and\ \bibinfo {author} {\bibfnamefont {I.}~\bibnamefont {Vega}},\ }\bibfield
  {title} {\bibinfo {title} {{The Motion of point particles in curved
  spacetime}},\ }\href {https://doi.org/10.12942/lrr-2011-7} {\bibfield
  {journal} {\bibinfo  {journal} {Living Rev. Rel.}\ }\textbf {\bibinfo
  {volume} {14}},\ \bibinfo {pages} {7} (\bibinfo {year} {2011})},\ \Eprint
  {https://arxiv.org/abs/1102.0529} {arXiv:1102.0529 [gr-qc]} \BibitemShut
  {NoStop}%
%%CITATION = ARXIV:1102.0529;%%
\bibitem [{\citenamefont {Gorbar}\ \emph {et~al.}(2011)\citenamefont {Gorbar},
  \citenamefont {Miransky},\ and\ \citenamefont {Shovkovy}}]{Gorbar:2011ya}%
  \BibitemOpen
  \bibfield  {author} {\bibinfo {author} {\bibfnamefont {E.~V.}\ \bibnamefont
  {Gorbar}}, \bibinfo {author} {\bibfnamefont {V.~A.}\ \bibnamefont
  {Miransky}},\ and\ \bibinfo {author} {\bibfnamefont {I.~A.}\ \bibnamefont
  {Shovkovy}},\ }\bibfield  {title} {\bibinfo {title} {{Normal ground state of
  dense relativistic matter in a magnetic field}},\ }\href
  {https://doi.org/10.1103/PhysRevD.83.085003} {\bibfield  {journal} {\bibinfo
  {journal} {Phys. Rev.}\ }\textbf {\bibinfo {volume} {D83}},\ \bibinfo {pages}
  {085003} (\bibinfo {year} {2011})},\ \Eprint
  {https://arxiv.org/abs/1101.4954} {arXiv:1101.4954 [hep-ph]} \BibitemShut
  {NoStop}%
%%CITATION = ARXIV:1101.4954;%%
\bibitem [{\citenamefont {Cao}\ and\ \citenamefont {He}(2019)}]{Cao:2019ctl}%
  \BibitemOpen
  \bibfield  {author} {\bibinfo {author} {\bibfnamefont {G.}~\bibnamefont
  {Cao}}\ and\ \bibinfo {author} {\bibfnamefont {L.}~\bibnamefont {He}},\
  }\bibfield  {title} {\bibinfo {title} {{Rotation induced charged pion
  condensation in a strong magnetic field: A Nambu\textendash{}Jona-Lasino
  model study}},\ }\href {https://doi.org/10.1103/PhysRevD.100.094015}
  {\bibfield  {journal} {\bibinfo  {journal} {Phys. Rev. D}\ }\textbf {\bibinfo
  {volume} {100}},\ \bibinfo {pages} {094015} (\bibinfo {year} {2019})},\
  \Eprint {https://arxiv.org/abs/1910.02728} {arXiv:1910.02728 [nucl-th]}
  \BibitemShut {NoStop}%
\bibitem [{\citenamefont {Frohlich}\ \emph {et~al.}(1981)\citenamefont
  {Frohlich}, \citenamefont {Morchio},\ and\ \citenamefont
  {Strocchi}}]{Frohlich:1981yi}%
  \BibitemOpen
  \bibfield  {author} {\bibinfo {author} {\bibfnamefont {J.}~\bibnamefont
  {Frohlich}}, \bibinfo {author} {\bibfnamefont {G.}~\bibnamefont {Morchio}},\
  and\ \bibinfo {author} {\bibfnamefont {F.}~\bibnamefont {Strocchi}},\
  }\bibfield  {title} {\bibinfo {title} {{Higgs phenomenon without symmetry
  breaking order parameter}},\ }\href
  {https://doi.org/10.1016/0550-3213(81)90448-X} {\bibfield  {journal}
  {\bibinfo  {journal} {Nucl. Phys.}\ }\textbf {\bibinfo {volume} {B190}},\
  \bibinfo {pages} {553} (\bibinfo {year} {1981})}\BibitemShut {NoStop}%
%%CITATION = NUPHA,B190,553;%%
\bibitem [{\citenamefont {Bricmont}\ and\ \citenamefont
  {Frohlich}(1983)}]{Bricmont:1983pq}%
  \BibitemOpen
  \bibfield  {author} {\bibinfo {author} {\bibfnamefont {J.}~\bibnamefont
  {Bricmont}}\ and\ \bibinfo {author} {\bibfnamefont {J.}~\bibnamefont
  {Frohlich}},\ }\bibfield  {title} {\bibinfo {title} {{An order parameter
  distinguishing between different phases of lattice gauge theories with matter
  fields}},\ }\href {https://doi.org/10.1016/0370-2693(83)91171-1} {\bibfield
  {journal} {\bibinfo  {journal} {Phys. Lett.}\ }\textbf {\bibinfo {volume}
  {122B}},\ \bibinfo {pages} {73} (\bibinfo {year} {1983})}\BibitemShut
  {NoStop}%
%%CITATION = PHLTA,122B,73;%%
\bibitem [{\citenamefont {Cao}\ and\ \citenamefont
  {Zhuang}(2015)}]{Cao:2015xja}%
  \BibitemOpen
  \bibfield  {author} {\bibinfo {author} {\bibfnamefont {G.}~\bibnamefont
  {Cao}}\ and\ \bibinfo {author} {\bibfnamefont {P.}~\bibnamefont {Zhuang}},\
  }\bibfield  {title} {\bibinfo {title} {{Effects of chiral imbalance and
  magnetic field on pion superfluidity and color superconductivity}},\ }\href
  {https://doi.org/10.1103/PhysRevD.92.105030} {\bibfield  {journal} {\bibinfo
  {journal} {Phys. Rev.}\ }\textbf {\bibinfo {volume} {D92}},\ \bibinfo {pages}
  {105030} (\bibinfo {year} {2015})},\ \Eprint
  {https://arxiv.org/abs/1505.05307} {arXiv:1505.05307 [nucl-th]} \BibitemShut
  {NoStop}%
%%CITATION = ARXIV:1505.05307;%%
\bibitem [{\citenamefont {Adhikari}(2019)}]{Adhikari:2018fwm}%
  \BibitemOpen
  \bibfield  {author} {\bibinfo {author} {\bibfnamefont {P.}~\bibnamefont
  {Adhikari}},\ }\bibfield  {title} {\bibinfo {title} {{Magnetic Vortex
  Lattices in Finite Isospin Chiral Perturbation Theory}},\ }\href
  {https://doi.org/10.1016/j.physletb.2019.01.027} {\bibfield  {journal}
  {\bibinfo  {journal} {Phys. Lett.}\ }\textbf {\bibinfo {volume} {B790}},\
  \bibinfo {pages} {211} (\bibinfo {year} {2019})},\ \Eprint
  {https://arxiv.org/abs/1810.03663} {arXiv:1810.03663 [nucl-th]} \BibitemShut
  {NoStop}%
%%CITATION = ARXIV:1810.03663;%%
\bibitem [{\citenamefont {Gell-Mann}\ \emph {et~al.}(1968)\citenamefont
  {Gell-Mann}, \citenamefont {Oakes},\ and\ \citenamefont
  {Renner}}]{GellMann:1968rz}%
  \BibitemOpen
  \bibfield  {author} {\bibinfo {author} {\bibfnamefont {M.}~\bibnamefont
  {Gell-Mann}}, \bibinfo {author} {\bibfnamefont {R.~J.}\ \bibnamefont
  {Oakes}},\ and\ \bibinfo {author} {\bibfnamefont {B.}~\bibnamefont
  {Renner}},\ }\bibfield  {title} {\bibinfo {title} {{Behavior of current
  divergences under SU(3) x SU(3)}},\ }\href
  {https://doi.org/10.1103/PhysRev.175.2195} {\bibfield  {journal} {\bibinfo
  {journal} {Phys. Rev.}\ }\textbf {\bibinfo {volume} {175}},\ \bibinfo {pages}
  {2195} (\bibinfo {year} {1968})}\BibitemShut {NoStop}%
%%CITATION = PHRVA,175,2195;%%
\bibitem [{\citenamefont {Yamamoto}(2014)}]{Yamamoto:2014lia}%
  \BibitemOpen
  \bibfield  {author} {\bibinfo {author} {\bibfnamefont {A.}~\bibnamefont
  {Yamamoto}},\ }\bibfield  {title} {\bibinfo {title} {{Lattice QCD with
  mismatched Fermi surfaces}},\ }\href
  {https://doi.org/10.1103/PhysRevLett.112.162002} {\bibfield  {journal}
  {\bibinfo  {journal} {Phys. Rev. Lett.}\ }\textbf {\bibinfo {volume} {112}},\
  \bibinfo {pages} {162002} (\bibinfo {year} {2014})},\ \Eprint
  {https://arxiv.org/abs/1402.3049} {arXiv:1402.3049 [hep-lat]} \BibitemShut
  {NoStop}%
\bibitem [{\citenamefont {Brauner}\ and\ \citenamefont
  {Huang}(2016)}]{Brauner:2016lkh}%
  \BibitemOpen
  \bibfield  {author} {\bibinfo {author} {\bibfnamefont {T.}~\bibnamefont
  {Brauner}}\ and\ \bibinfo {author} {\bibfnamefont {X.-G.}\ \bibnamefont
  {Huang}},\ }\bibfield  {title} {\bibinfo {title} {{Vector meson condensation
  in a pion superfluid}},\ }\href {https://doi.org/10.1103/PhysRevD.94.094003}
  {\bibfield  {journal} {\bibinfo  {journal} {Phys. Rev.}\ }\textbf {\bibinfo
  {volume} {D94}},\ \bibinfo {pages} {094003} (\bibinfo {year} {2016})},\
  \Eprint {https://arxiv.org/abs/1610.00426} {arXiv:1610.00426 [hep-ph]}
  \BibitemShut {NoStop}%
%%CITATION = ARXIV:1610.00426;%%
\end{thebibliography}%

\end{document}